\documentclass[prb,aps,showpacs,superscriptaddress,twocolumn]{revtex4}
\usepackage{color} 
\usepackage{times}
\usepackage{amssymb} 
\usepackage{amsmath} 
\usepackage{graphicx}
\usepackage{epstopdf}
\usepackage[english]{babel}
\usepackage{mathrsfs}
\usepackage{textcomp}
\usepackage{amsthm}
\usepackage{amsfonts}
\usepackage{amssymb} 
\usepackage{hyperref}
\usepackage{soul}

\newcommand{\mean}[2]{\langle#1\rangle_{#2}}

\newcommand{\opt}[4]{\hat{#1}_{#3}^{#4}(#2)}
\newcommand{\op}[3]{\hat{#1}_{#2}^{#3}}

\begin{document}

\title{Self-consistent Keldysh approach to quenches in weakly interacting Bose-Hubbard model}

\author{N. Lo Gullo}
\affiliation{Dipartimento di Fisica e Astronomia ``G. Galilei'', 
Universit\`a degli Studi di Padova, and CNISM,  
Padova, Italy}
\affiliation{Dipartimento di Fisica, Universit\`a degli Studi di Milano,
Milano, Italy}
\author{L. Dell'Anna}
\affiliation{Dipartimento di Fisica e Astronomia ``G. Galilei'', 
Universit\`a degli Studi di Padova, and CNISM, 
Padova, Italy}

\begin{abstract}
 We present a non-equilibrium Green's functional approach to study the 
 dynamics following a quench in weakly interacting Bose-Hubbard model (BHM).
 The technique is based on the self-consistent solution of a set of equations
 which represents a particular case of the most general set of Hedin's equations for the interacting
 single-particle Green's function. We use the ladder approximation as a skeleton diagram for the 
 two-particle scattering amplitude useful, through the self-energy in the 
 Dyson equation, for finding the interacting single-particle Green's function.
 This scheme is then implemented numerically by a parallelized code.
 We exploit this approach to study the correlation propagation after a 
 quench in the interaction parameter, for one (1D) and two (2D) dimensions.
 In particular, we show how our approach is able to recover the  
crossover from 
 ballistic to diffusive regime by increasing the boson-boson interaction.
 Finally we also discuss the role of a thermal initial state on 
the dynamics both for 1D and 2D BHMs, finding that 
surprisingly at high temperature a ballistic evolution is restored. 
\end{abstract}

\maketitle

\section{Introduction}
The study of out-of-equilibrium quantum systems has received quite some 
attention in recent years and many efforts have been devoted to understand 
properties of thermalization or relaxation towards equilibrium in the quantum 
regime \cite{silva2011,eisert2015,huse2015}, transport \cite{wang2006},
dynamical phase transitions \cite{fischer2006,fischer2008,kollath2007,altman2014,eisert2015}, 
ergodic and non-ergodic quantum systems \cite{huse2015}.
The quickly developing field of ultracold atomic gases offers
the possibility of manipulating and controlling complex quantum systems
with very high accuracy. This tunability has undoubtedly increased the 
interest in studying 
the dynamics in such systems due to the possibility of experimentally testing
theoretical predicitons and offering new research perspectives \cite{bloch2008,citro2011,bloch2012}. 
The Bose-Hubbard model (BHM) is one of the most studied system in this context,
because of its rich phase diagram \cite{fisher,kennett2013} 
and because it describes faithfully the dynamics of ultracold atomic bosons 
in optical lattices \cite{kollath2007,kennett2013,kennett2011,kennett2016}, 
and therefore it is relevant for experimental investigations.
In order to theoretically study the BHM generally one can resort to the following numerical tools: exact numerical diagonalization \cite{kollath2007,zhang2010} 
and time dependent density matrix renormalization group (t-DMRG) approach \cite{cazalilla2002}.

A very powerful method for non-equilibrium systems is the 
Keldysh-Green's functional approach developed to describe the dynamics of a 
quantum system without the assumption that the system relaxes towards a steady state \cite{keldysh1965,kamenev}.
This technique can be formulated
either in the form of equations of motion for the single-particle 
Green's function or
as a many-body perturbation approach,   
eventually resumming some of the diagrams appearing in the series.
For the single particle Green's function, and for a two-body interaction Hamiltonian,
it is possible to define a closed set of equations, whose iterative solution
is the solution to the initial problem \cite{stefanucci}.
This set of equations has been named after Hedin the Hedin's equations.
The iterative solution of such equations is obviously found by means
a proper (depending upon the problem at hand) numerical approach. 
Hedin's equations are nevertheless difficult to be implemented 
because of the appearance of a functional derivative in
the equation for the vertex function.
One would thus need to find a way to calculate it analytically at different 
orders and then insert it into the numerical iteration.
This process is nevertheless rather complicated and often the way out
is to drop this equation thus effectively keeping the bare vertex
at all iteration steps \cite{stefanucci}.

However it is possible to recast Hedin's equations 
in a different, but fully equivalent, set of equations 
as shown by Starke and Kresse in their work \cite{starke2012}.
The idea behind this is to exploit the fact that, for a two-body
interaction Hamiltonian, the time evolution of the 
interacting single-particle Green's function is related only 
to the two-body Green's function.
It is then possible to write a set of equations for both
single- and two- particle Green's functions and then
solve it iteratively.
In our work we apply a slightly different version of this idea to study the dynamics
of a BHM following a sudden quench in the boson-boson interaction strenght.
Our approach is fully consistent with the work of Starke and Kresse\cite{starke2012}
and the only difference is that we will use the two-particle scattering amplitude 
instead of the two-particle Green's function.

The remainder of the paper is organized as follows: in Sec. \ref{sec:selfmeth}
we describe the self-consistent methods and discuss their properties,
limitation and potentiality, in Sec. \ref{sec:bhm} we introduce the perturbative theory 
used to study the dynamics of the weakly interacting Bose Hubbard model (BHM),
whereas in Sec. \ref{sec:selfcons} we describe the corresponding iterative scheme used to
find the time-dependent single particle Green's functions and its relation 
to the previously introduced schemes.
Finally, in Secs. \ref{sec:1D} and \ref{sec:2D} we present a study of quenches 
in one and two dimensional BHM for different paramenters of the system such as 
initial chemical potential, final interaction, temperature, and dimensionality.

\section{Self-consistent methods}
\label{sec:selfmeth}
In this section we briefly review self-consistent methods
which allow one to calculate the single-particle Green's function 
for an interacting many-body system with a two-body interaction.
To fix the ideas we will consider from now a bosonic system described 
by the Hamiltonian:
\begin{eqnarray}
 \label{eq:fieldham}
 \mathcal{\hat{H}}&=&\mathcal{\hat{H}}_0+\mathcal{\hat{V}}(t)\\
 &&\nonumber\\
 \mathcal{\hat{H}}_0&=&-\frac{\hbar^2}{2m}\int\; d{\bf x}\;\opt{\phi}{{\bf x},t}{}{\dag}\Delta\opt{\phi}{{\bf x},t}{}{}\nonumber\\
 &+&\int d{\bf x} \;v({\bf x})\opt{\rho}{\bf x,t}{}{}\nonumber\\
 &&\nonumber\\
 \mathcal{\hat{V}}(t)&=&\frac{1}{2}\int d{\bf x}d{\bf x'}u(|{\bf x}-{\bf x'}|,t)\nonumber\\
 &\times&\opt{\phi}{\bf x,t}{}{\dag}\opt{\phi}{\bf x',t}{}{\dag}\opt{\phi}{\bf x',t}{}{}\opt{\phi}{\bf x,t}{}{}.\nonumber
\end{eqnarray}
where $\opt{\phi}{{\bf x},t}{}{}$ and $\opt{\phi}{{\bf x},t}{}{\dag}$ are bosonic field operators
satisfying commutation relations $[\opt{\phi}{{\bf x},t}{}{},\opt{\phi}{{\bf x}',t'}{}{\dag}]=-\imath\hbar\delta({\bf x}-{\bf x'})\delta(t-t')$ 
and $[\opt{\phi}{{\bf x},t}{}{},\opt{\phi}{{\bf x}',t'}{}{}]=0$.
We defined $\opt{\rho}{{\bf x},t}{}{}=\opt{\phi}{{\bf x},t}{}{\dag}\opt{\phi}{{\bf x},t}{}{}$
and allowed for an explicit time dependence of the interaction Hamiltonian.
Most of the interesting information about the system 
such as density, spectra, response functions,
can be extracted from the knowledge of the single-particle Green's function
defined as $G(1;1')=-\imath\;\left\langle \mathcal{T}_\gamma\opt{\phi}{1}{}{}\opt{\phi}{1'}{}{\dag} \right\rangle
$, where $\gamma$ is the (properly chosen) Keldysh contour \cite{keldysh1965,kamenev,stefanucci}, 
$\mathcal{T}_\gamma$ is the time ordering over $\gamma$
and we used the standard notation for variables, 
namely $1=\{{\bf x}_1,z_1\}$ and $1'=\{{\bf x}_1',z_1'\}$,
$z$ being the complex time variable on the contour.
By properly choosing the contour $\gamma$ different approaches can be recovered~\cite{stefanucci}:
for instance by choosing the contour to run only along the imaginary axis
the Matsubara approach to study thermodynamic properties of interacting systems
is obtained.
We will focus on a contour which goes around the real line running from 
$z_0^{+}=t_0+\imath \epsilon$ to $z_0^{-}=t_0-\imath \epsilon$ $(\epsilon>0)$ 
crossing the real line at $z_f=t_f$, namely the maximum evolution time. 

As clearly described in Ref.~\cite{starke2012}, 
in order to find the Green's function one can resort to two different, but equivalent approaches.
The first uses the Heisenberg equations for the field operators to find an equation
of motion for the Green's function. This method results in a hierarchy
of equations, the Martin-Swinger hierarchy, in which Green's function 
of different orders are linked to each other.
In our case (two body interaction Hamiltonian) the hierarchy links
the time derivative of the $n-$particle Green's function to the 
$n+1-$particle Green's function. So that the single-particle Green's functions is 
related to the two-particle one.
The second approach is based on the Gell-Mann Low theorem,
allowing the cancellation of disconnected diagrams (linked cluster expansion),
in order to calculate the Green's function perturbatively by including 
the wanted set of diagrams whose choice depends upon the physics of the system at hand. 
This technique is particularly powerful because 
it allows us to take into account the chosen diagrams to all order in the 
coupling constant, thus representing an evolution of simple perturbative approach.
Here we will follow this second approach.

In this case it is possible to derive a closed set of equations,
the Hedin's equations, whose (iterative) solution gives 
the interacting single-particle Green's function:
\begin{eqnarray}
 &&G(1;1')=G_0(1;1')+\!\!\int \!\!d\overline{1} d\overline{2}\;G_0(1;\overline{1})\Sigma(\overline{1};\overline{2}) G(\overline{2};1')\\
 \label{eq:dyson}
 &&\Sigma(1;2)=\Sigma_H[G,u](1;2)+\Sigma_{ex}[G,W](1;2)\\
 \label{eq:selfen}
 &&P(1;2)=\imath \int d\overline{1} d\overline{2}\; G(1;\overline 1) G(\overline 2;1^+) \Lambda(\overline 1;\overline 2;2)\\
 \label{eq:pol}
 &&W(1;2)=u(1;2)+\!\!\int\!\! d\overline{1} d\overline{2}\;u(1;\overline{1})P(\overline{1};\overline{2})W(\overline{2};2)\\
 \label{eq:dressint}
 &&\Lambda(1,3;2)=\delta(1,3^{+})\delta(2,3)\\
 &&+\imath\!\!\int \!\! d\overline{1} d\overline{2}d\overline{3} d\overline{4}\;\frac{\delta \Sigma_{ex}[G,W](1,3)}{\delta G(\overline 1;\overline 2)}G(\overline 1;\overline 3) G(\overline 4;\overline 2)\Lambda(\overline 3, \overline 4;2)\nonumber
\label{eq:vert}
\end{eqnarray}
where $u(1;2)=u(|{\bf x}_1-{\bf x}_2|,z_1)\delta_\gamma(z_1-z_2)$.
These equations are: the Dyson equation for the single particle Green's function;
the equation for the self-energy where $\Sigma_H$ and $\Sigma_{ex}$ 
are the Hartree and exchange self-energy respectively;
the equation for the polarization; the equation
for the dressed interaction; the equation for the vertex function.
All these equations have to be solved iteratively with properly chosen seeds for the iterative scheme;
for instance the initial single-particle Green's function is often chosen to be either 
the non-interacting one or the self-consistent Hartree Green's function.
Nevertheless in trying to solve Hedin's equations iteratively  
different problems arise.
The most difficult to overcome is related to the fact that
the equation for the vertex function contains a functional derivative of the 
(exchange) self-energy with respect to the interacting single particle Green's function.
This makes the numerical implementation particularly difficult.
Schindlmayr and Godby ~\cite{godby1998} have obtained an analytical expression for the vertex function
after the first iteration step for the Fermi-Hubbard model.
Nevertheless Hedin's approach in its generality still suffers from this problem
and in most cases of interest implementing such a scheme is so complicated 
that other options are usually preferred.

Recently Starke and Kresse used a slightly different point of view in the 
derivation of a self-consistent set of equations \cite{starke2012}.
After observing that, as mentioned above,
the single-particle Green's function for a two-body interaction Hamiltonian,
is related to the two particle one,
they derived a set of two equations 
which is equivalent to Hedin's equations:
\begin{eqnarray}
 &&G(1;1')=G_0(1;1')\\
 &&+\imath\int \!\!d\overline{1} d\overline{2}\; G_0(1;\overline{1})u(\overline{1};\overline{2}) L(\overline{2}, \overline{2};\overline{1}, 1')\nonumber\\
 \label{eq:greensk}
 &&L(1,2;1',2')=L_0(1,2;1',2')\\
 &&+\imath\int \!\!d\overline{1} d\overline{2}d\overline{3} d\overline{4}\;
L_0(1,\overline{1};1',\overline{2})I(\overline{2},\overline{3};\overline{1},\overline{4})L(\overline{4},2;\overline{3},2')\nonumber
\label{eq:vertsk}
\end{eqnarray}
where $L(1,2;1',2')=G_2(1,2;1',2')-G(1;1')G(2;2')$
with $G_2$ being the interacting two-particle Green's function.
The two equations are: 1) an equation for the interacting single-particle Green's function
(analogous to the Dyson equation, Eq.~\ref{eq:dyson});
2) the Bethe-Salpeter equation for the exchange part of the interacting 
two-particle Green's functions.
This set of equations does not rely on any functional derivative,
but it fixes {\it a priori} the functional form of the 
kernel $I$ of the Bethe-Salpeter equation,
namely the two-particle scattering amplitude, which on the other hand can be 
upgraded at every iteration depending on the choice of both $I$ and $L_0$ \cite{starke2012}.
The latter is in general a functional of the {\it interacting} single particle
Green's function as well.
Depending on the choice of $I$ and $L_0$ it is possible 
to obtain different schemes, including the most used in literature such as
the Hartree-fock approximation, 
the Random-Phase approximation, the GW-approximation,
the ladder approximation and the W-approximation
(see Table I in Ref.~\cite{starke2012}
for the hierarchy of approximation for different choices of the kernel $I$).

In the following we will introduce our approach closely related 
to the one derived by Starke and Kresse to find self-consistently 
the one- and two-particle Green's functions for the weakly interacting
Bose-Hubbard model.

\section{Weakly interacting Bose-Hubbard model}
\label{sec:bhm}
The Bose Hubbard model (BHM) is obtained as the tight-binding limit of
the Hamiltonian in Eq.\ref{eq:fieldham} where the potential is chosen to have 
several local minima.
In this limit we can choose a basis of wavefunctions
localized around the local minima of the potential and 
write the field operator as $\opt{\phi}{{\bf x},t}{}{}=\sum_i w_i({\bf x})\opt{b}{t}{i}{}$.
The Hamiltonian thus reads
\begin{eqnarray}
\op{H}{}{}=&&\op{H}{0}{}+\opt{V}{t}{}{}\\
\op{H}{0}{}=&&\sum_{i}\epsilon_i\;\op{b}{i}{\dag}\op{b}{i}{}-\sum_{<i,j>}\frac{J}{2}\left(\op{b}{i}{\dag}\op{b}{j}{}+\text{h.c.}\right)\\
\opt{V}{t}{}{}=&&\frac{U(t)}{2}\;\sum_{i}\op{b}{i}{\dag}\op{b}{i}{\dag}\op{b}{i}{}\op{b}{i}{},
\label{eq:totham}
\end{eqnarray}
The operators $\op{b}{i}{}{}$ and $\op{b}{i}{\dag}{}$ obey bosonic commutation
relations: $\left[\op{b}{i}{}{},\op{b}{j}{\dag}{}\right]=\delta_{i,j}$
and $\left[\op{b}{i}{}{},\op{b}{j}{}{}\right]=0$.
To fix the ideas we will think of interacting bosonic atoms in an optical lattice
for which the BHM has been proven to give a very good description \cite{kennett2013}.
In what follows we shall consider the weakly interacting limit $U(t)<J,\;\forall t$ 
and resort to a perturbative expansion of the single particle 
interacting Green's functions.
The latter in the interaction picture reads
\begin{equation}
 G(1,1')=-\imath\;\frac{\left\langle \mathcal{T}_\gamma\left[e^{-\imath\int_{\gamma}dz \opt{V}{z}{I}{}}\opt{b}{1}{}{}\opt{b}{1'}{}{\dag}\right] \right\rangle
}{\left\langle \mathcal{T}_\gamma\left[ e^{-\imath\int_{\gamma}dz \opt{V}{z}{I}{}}\right]\right\rangle},
\label{eq:onebodyG}
\end{equation}
where $\opt{V}{z}{I}{}=U(z)/2\; \sum\limits_{i}\opt{b}{z^{+}}{i}{\dag}\opt{b}{z^{+}}{i}{\dag}\opt{b}{z}{i}{}\opt{b}{z}{i}{}$
is the interaction Hamiltonian in the interaction picture and the multi-index
is now $1=\{i_1,z_1\}$, $i_1$ being the site index and $z_1$ the complex time on the contour.
In order to calculate $G(1,1')$ we resort to its expansion in terms of 
the non-interacting Green's functions $g(1,1')$.
In particular we use the so called ladder approximation,
which amounts to calculate the self-energy entering the Dyson equation
by means of a two-particle scattering amplitude which in turn is calculated 
from the Bethe-Salpeter equation.
For the interaction term as in Eq.\ref{eq:totham},  
$G(1;1')$ can be approximated to second order
by expanding the evolution operator. 
Assuming the initial state is such that the non-interacting $n$-particle
Green's functions can be written as a permanent 
of the non-interacting single-particle ones
the denominator in Eq. \ref{eq:onebodyG} 
cancles all disconnected diagrams in the expansion
thus leaving only connected ones \cite{stefanucci}.

In order to construct our iterative method let us look at
the first and second order (connected) contributions which are given by:
\begin{eqnarray}
\label{eq:fsorders}
 G_c^{(1)}(1;1')&=&2 \imath\int_\gamma d\overline 1\;
 U(\overline z_1) g(1;\overline 1^+)g(\overline 1;\overline 1^+)g(\overline 1;1')\\
 G_c^{(2)}(1;1')&=&2\;\imath^2\int_\gamma d\overline 1 d\overline 2\;
 U(\overline z_1)U(\overline z_2)\;\\
 &&g(1;\overline 1^+)g(\overline 1;\overline 2^+)g(\overline 2;\overline 1^+)g(\overline 1;\overline 2^+)g(\overline 2;1')\nonumber
\end{eqnarray}
In the second order we considered only the two-particle irreducible contributions, 
{\it i.e.} those with no self-energy insertions, the reason will become clear
in the following.
The factor two comes from the fact that direct and exchange integrals 
give the same contribution due to the on-site nature of the interaction
and to the lack of any other label but the site index of the bosonic operators.

The above terms can be recovered from the Dyson equation
for the interacting single-particle Green's function with 
the definition of a self-energy.
We are going to define it by means of a vertex function 
which is in turn recovered from the Bethe-Salpeter equation 
whose kernel we chose to be the bare interaction:
\begin{eqnarray}
 &&K(1',2';1,2)=V(1',2';1,2)\\
 &&+\imath \int d \overline{1}d \overline{2}d \overline{3}d 
\overline{4}\;V(1',2';\overline{1},\overline{2})g(\overline{1};\overline{3}
)g(\overline{2};\overline{4})K(\overline{3},\overline{4};1,2)\nonumber
 \label{eq:KBS}
\end{eqnarray}
where in our case $V(1',2';1,2)=U(z_1)\delta(2-1)\delta(1'-1^+)\delta(2'-2^+)$
and $\delta(1'-1^+)=\delta_{i_1' i_1}\delta_\gamma(z_1'-z_1^{+})$
and we assumed an interaction Hamiltonian of the form 
$V_I=V(1',2';1,2)=\opt{b}{1'}{}{\dag}\opt{b}{2'}{}{\dag}\opt{b}{1}{}{}\opt{b}{2}{}{}$.
We then define the self-energy as
\begin{equation}
 \Sigma(1';1)=\imath\int d\overline{1}d\overline{2}\;(K(1',\overline 
2';1,\overline{2})+K(1',\overline 
2';\overline{2},1))g(\overline{2};\overline{2}').
 \label{eq:SigmaBS}
\end{equation}
This is known as the ladder approximation 
from the form of Feynman diagrams at different orders for the two-particle
scattering amplitude $K$.
In the case of interacting electrons the ladder approximation is associated to
the low density limit and it is physically justified 
because the contribution of particle-hole like excitation is of higher order (in the interaction) 
with respect to particle-particle scattering in this limit \cite{mattuck,fettwal}.
On the other hand it can also be used to study dilute bosonic gases and their thermodynamical properties\cite{fettwal}.

The definition of a self-energy allows us to write the Dyson equation
for the interacting single-particle Green's function:
\begin{equation}
 G(1;1')=g(1;1')+\int d\overline{1}d\overline{1}'\;g(1;\overline{1'})\Sigma(\overline{1}',\overline{1}) G(\overline{1};1').
 \label{eq:Dyson}
\end{equation}

\section{Self-consistent solution and its numerical implementation}
\label{sec:selfcons}
We have seen that through the definition of a (proper) self-energy 
it is possible to write a Dyson equation for the interacting single-particle 
Green's function.
Nevertheless the self-energy defined above
does not allow to recover many important contributions,
in particular those which are one-interaction line reducible.
This is because the two-particle scattering amplitude $K$,
and therefore the self-energy $\Sigma$ derived from it,
does not contain such diagrams by construction.
For instance, the following second order contribution
cannot be obtained from the Dyson equation with the self-energy
defined in Eq.~(\ref{eq:SigmaBS}), 
$ \imath^2\int d\overline{1} d\overline{2}\,
 U(\overline z_1)U(\overline z_2)g(1;\overline 1)g(\overline1;\overline 2^{+}) 
g(\overline 2;\overline 2^{+})g(\overline2;\overline 1^{+})g(\overline 1;1')$. 
The above second order term can be accounted for
by using the Hartree Green's function
in the Dyson equation instead of the non-interacting one.
Nevertheless there are higher order contributions 
which cannot be derived by means of the 
Hartree propagator and the ladder approximation 
and they would require more complex self-energies to be defined.
To get around this problem we resort to iterations 
by means of the following iterative scheme:
\begin{widetext}
\begin{eqnarray}
\label{eq:iteration}
K_{i_1j_1i_2j_2}^{(n)}(z_1;z_2)&=&V_{i_1j_1i_2j_2}(z_1;z_2)\nonumber\\
 &+&\imath\sum_{\substack{\overline{i}_1,\overline{j}_1\\ \overline{i}_2,\overline{j}_2}}\int_{\gamma}d \overline{z}_1 d \overline{z}_2\; V_{i_1 j_1\overline{i}_2 \overline{j}_2}(z_1;\overline{z}_2)\;G_{\overline{i}_2\overline{i}_1}^{(n-1)}(\overline{z}_2;\overline{z}_1)\;G_{\overline{j}_2\overline{j}_1}^{(n-1)}(\overline{z}_2;\overline{z}_1)\;K_{\overline{i}_1 \overline{j}_1i_2 j_2}^{(n)}(\overline{z}_1;z_2)
 \label{eq:itK}\\
 {\Sigma}_{i_1i_2}^{(n)}(z_1;z_2)&=&\imath\sum_{\overline{i}_1,\overline{i}_2} {K}_{i_1 \overline{i}_1i_2 \overline{i}_2}^{(n)}(z_1;z_2)G_{\overline{i}_2\overline{i}_1}^{(n-1)}(z_2;z_1)+\imath\sum_{\overline{i}_1,\overline{i}_2} {K}_{i_1 \overline{i}_1 \overline{i}_2i_2}^{(n)}(z_1;z_2)G_{\overline{i}_2\overline{i}_1}^{(n-1)}(z_2;z_1)
 \label{eq:itsigma}\\
{G}_{i_1i_2}^{(n)}(z_1;z_2)&=&G_{i_1 i_2}^{(0)}(z_1;z_2) + \sum_{\overline{i}_1,\overline{i}_2}\int_{\gamma}d\overline{z}_1 d\overline{z}_2\; G_{i_1\overline{i}_1}^{(0)}(z_1;\overline{z}_1)\Sigma_{\overline{i}_1\overline{i}_2}^{(n)}(\overline{z}_1;\overline{z}_2){G}_{\overline{i}_2 i_2}^{(n+1)}(\overline{z}_2; z_2) 
\label{eq:itdyson},
\end{eqnarray}
\end{widetext}
with initial seed ${G}_{i_1 i_2}^{(0)}(z_1 z_2)={g}_{i_1 i_2}(z_1 z_2)$
being the non-interacting Green's functions and $n$ being the iteration step.
In the above equations we split the time and the site indeces, 
which will turn useful when rewriting the equation for the different 
components on the contour (see Appendix \ref{app:realtime}).
This is nothing but a subset of the most general system of 
equation, namely the Hedin's equations.
It is possible to show that the above set of equations is equivalent to the 
one in Eqs.~(\ref{eq:greensk}) 
with the choice $I(1,2;3,4)=V(1,2;3,4)+V(1,2;4,3)$
and $L_0(1,2;3,4)=G(1;3)G(2;4)$ \cite{starke2012}.
The difference between our approach and the one presented in 
Ref.~\cite{starke2012} 
relies on the fact that rather than working with the two-particle Green's 
functions (namely $L$) we use the two-particle scattering amplitude ($K$).
The latter is a tensor which is partially diagonal (Appendix \ref{app:realtime}) 
and can thus be implemented as a matrix in the numerical simulations. 
The two particle Green's function requires four (spatial) indexes in general
and it would be more costly in terms of computational resourses.
It is important to notice that the self-consistent ladder approximation is 
a conservative scheme in the sense of Kadanoff and Baym\cite{kadbay1961}.

In order to make the above system of equations suitable for numerical 
implementation, each equation has to be decomposed into the corresponding equations 
with real time arguments which can be done by means of Langreth theorem.
The result is the set of equations presented in Appendix \ref{app:realtime}.
The numerical implementation is then performed by means of two Fortran 90 codes,
the first of which solves the non-interacting problem relative to $\hat{H}_0$
and passes the non-interacting Green's functions $g$ to the second code 
which then implements the above iterative scheme.
The first code uses OpenMP to speed up the computation of
the non-interacting Green's function. 
The second code uses MPI parallelization by means 
of the ScaLAPACK libraries on a grid of (usually) $128$ processes.

In what follows we will use the iterative scheme in 
Eqs.~(\ref{eq:itK})-(\ref{eq:itdyson})
to study the dynamics of one- and two-dimensional Bose-Hubbard model
following the switching on of the boson-boson interaction $U$.
We also compared our approach with results from exact numerical diagonalization (see Appendix \ref{app:exact})
finding good agreement and confirming that we can rely on it for the range of parameter considered
in this work.

\section{Figures of merit}
To characterize the post-quench dynamics of the BHM we will look at both the 
expansion of the density of bosons and the spreding of correlations
The spreading of bosons
over the lattice will be characterized by means of the time-dependent standard deviation
of the normalized distribution obtained as $p_i(t)=n_i(t)/n$ ($\;\sum_i p_i(t)=1\;\;\;\forall t$):
$\sigma(t)=(\overline{i^2}(t)-\overline{i}(t)^2)^{1/2}$ where $\overline{i}(t)=\sum_i i p_i(t)$
and $\overline{i^2}(t)=\sum_i i^2 p_i(t)$
are the average with respect to the probability distribution $p_i(t)$
of the position and its square.
The velocity of propagation is given by the time derivative 
of $v(t)=d\sigma(t)/dt$.
We will also look at the momentum distribution defined through the single particle density matrix:
\begin{eqnarray}
n(k)=\frac{\imath}{\sqrt{N}}\sum\limits_{i} e^{-\imath k(i-j)} G_{i,j}^{<}(t;t^{+}).
\end{eqnarray} 

In order to characterize the spreading of correlations we will look at the evolution of the 
particle propagator and in particular at its variation $|\delta G_{ii_0}^{>}(t;0)|^2=|G_{ii_0}^{>}(t;0)-G_{ii_0}^{>}(0;0)|^2$,
between the site $i_0$ and all other sites as a function of time.
$G_{ii_0}^{>}(t;0)$ gives information on the mobility of a single-particle through the system.
In the remainder of the paper we will fix $i_0=\lfloor N/2\rfloor$,
other choices changes the results only quantitatively but not qualitatively.
The importance of this quantity (and in general of two-times ones) 
is related to its r\^ole in the determination
of the response of observables of the system to external perturbations,
in our case to the quench in the interaction.

\section{Quenches in 1D Bose-Hubbard model}
\label{sec:1D}
In this section we exploit the presented approach to study 
the dynamics of a homogeneous one-dimensional BHM following a quench 
in the boson-boson interaction.
In a recent work \cite{logullo2015} we have shown that the speed of propagation
of correlations depends upon the spectrum of the final Hamiltonian.
Specifically the speed increases by increasing the boson-boson interaction
as suggested by the dispersion relation of excitations 
in the Bogoliubov theory in the presence of repulsive interaction.
Our analysis was limited to small interaction and relied on the Bogoliubov
approach, thus giving an effectively non-interacting theory.
This is why we could access the ballistic regime only.
However it is expected that by increasing the boson-boson interaction
the system eventually enters a diffusive regime due to the 
non-linearity introduced by interactions; 
this behavior has also been observed experimentally \cite{ronzheimer2013}
showing that, inside the superfluid phase, as the final interaction strenght is
increased, the propagation of particles (density) is diffusive
unlike at small interaction where it is ballistic.
We now show that the approach introduced above is able to 
capture such a feature.

\subsection{Zero temperature}
\begin{figure}[h]
\includegraphics[width=8cm]{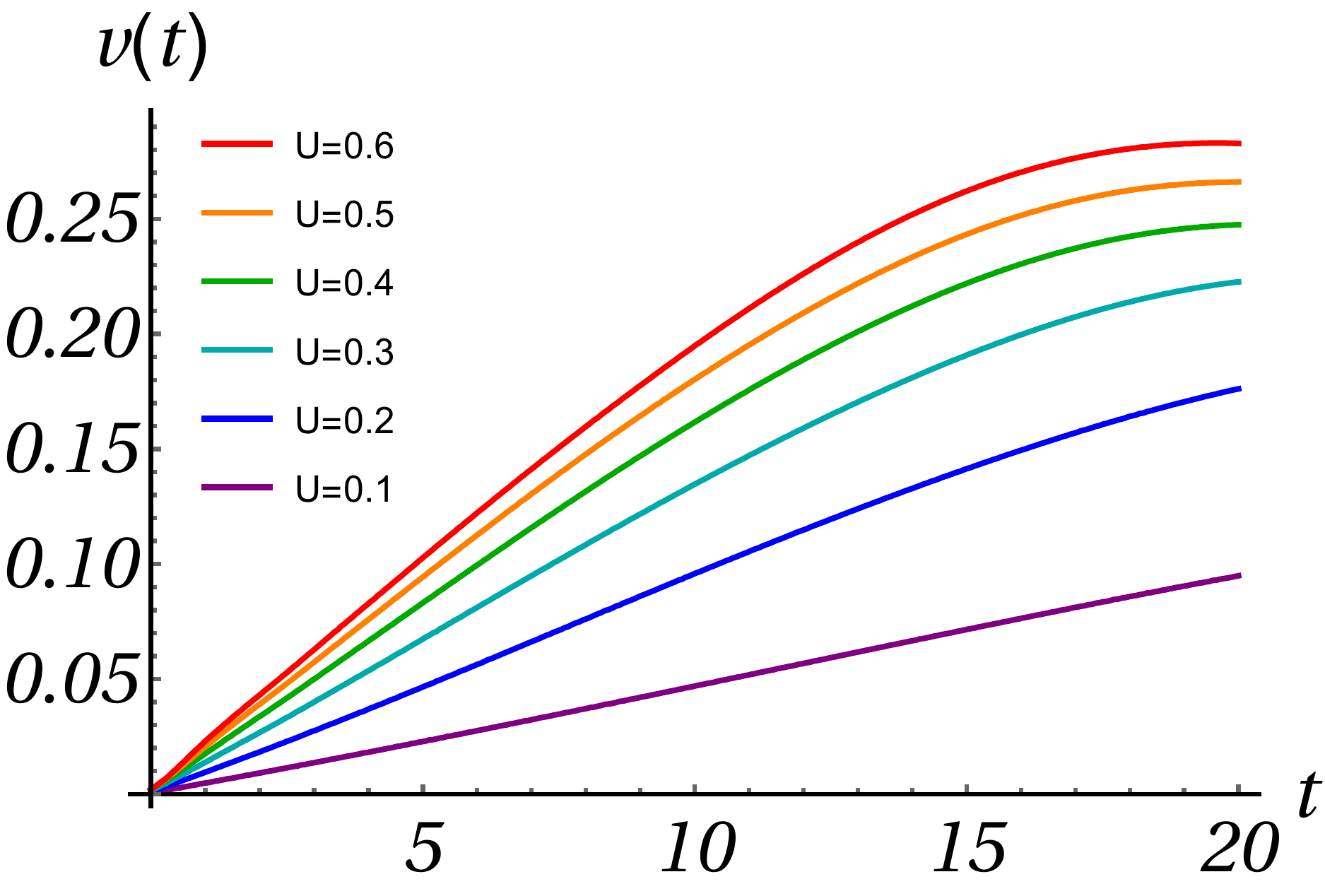}
\caption{(Color online). Velocity $v(t)$ (see text) for a system with $N=63$ sites 
and $n=21$ particles initially in the ground state of the (non-interacting) system.
Different plots are for different final boson-boson interactions. We set $J=1$.}
 \label{fig:1Dvt}
\end{figure}

\begin{figure}[h]
\includegraphics[width=8cm]{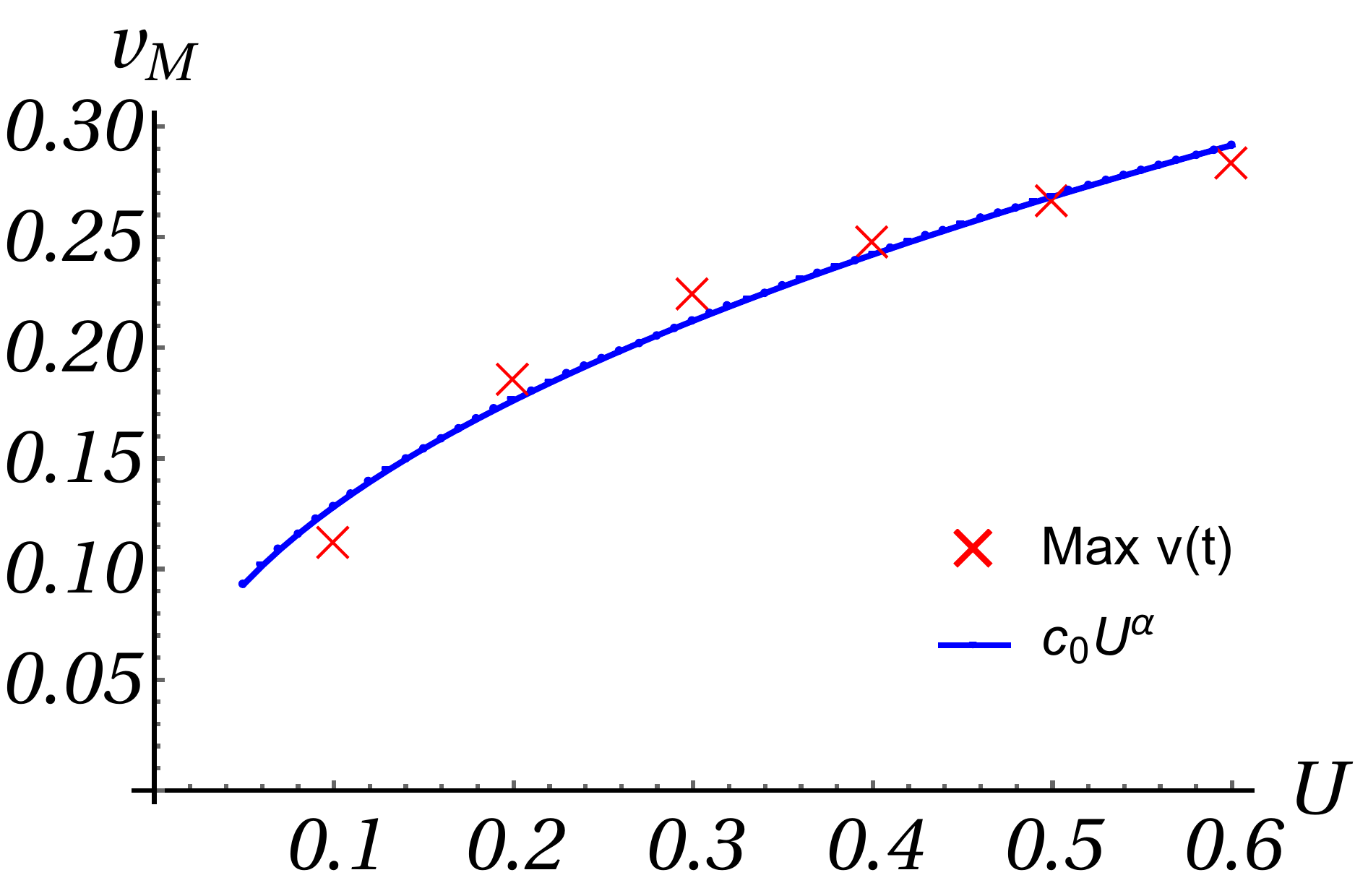}
\caption{(Color online). Maximum velocity as a function of $U$ obtained from the 
fit of the curves in Fig.\ref{fig:1Dvt} with a polynomial of sixth order in 
$t$. 
The fitting curve is obtained as $v_{M}(U)=c_0 U^\alpha$ with $c_0=(0.368 \pm 0.018)$
and $\alpha=(0.459\pm 0.047)$.}
 \label{fig:1Dvmax}
\end{figure}

We start by looking at the expansion of the boson density as a function of time and in particular 
at the speed of propagation $v(t)$ defined above.
In Fig.\ref{fig:1Dvt} we show $v(t)$ for a system with $N=63$ sites and 
$n=21$ bosons for quenches in the interaction from an initially non-interacting state.
We can see that the speed increases in time thus showing 
the accelerated character of the expansion of the density.
This was obviuosly to be expected since the sudden switching-on of the interaction
corresponds to a sudden inhomogeneous increase of the energy across the system
and in particular it is more concentrated where the density of particles is higher
at $t=0$ (the center).
This interaction energy is readly converted into kinetic energy resulting in an
accelerated expansion of the density towards the boundaries.
We observe that for small interactions the velocity is smaller than
for higher ones and this is obviously due to the total initial energy provided during the quench.
On the other hand it is interesting to notice that the asymptotic value of the 
velocity is reached faster at higher interactions (red and orange curves).
This can be explained by the fact that although the energy provided to the system 
increases with the interaction, the rate at which this energy is converted into kinetic
energy and distributed across the system is obviously higher for higher interactions
due to the increase in the particle-particle scattering processes.
We can assume that there exist a transient time, before the system reaches
its stationary state, in which the expansion of the density is accelerated
up until some time $t^*(U)$ after which the expansion slows down and 
the system starts to equilibrate. This characteristic time corrensponds 
roughly to the time at which the expansion reaches its maximum speed.
From this argument we can also expect that the maximum speed reached at $t=t^*(U)$
is such that $v_{M}\propto \sqrt{U}$ because when the number of bosons per site 
will be $n_i\approx 1$ they will become effectively non-interacting and 
therefore all initial energy, which is $\propto U$, will be converted into kinetic energy.
We can check this statement by extrapolating the behavior of the velocity $v(t)$
at longer times. 
We used a polynomial of sixth order in $t$
to fit $v(t)$ for each value of $U$ and then we extrapolated the maximum of this function in $t$.
These values are plotted in Fig.\ref{fig:1Dvmax}.
We also show the fitting function of the maxima as a function of $U$
which gives $v_{\infty}(U)=c_0 U^\alpha$ with $c_0=(0.368 \pm 0.018)$
and $\alpha=(0.459\pm 0.047)$ and therefore it is in agreement with our
expectation $v_{M}(U)\propto \sqrt{U}$.

\begin{figure}
\includegraphics[width=8cm]{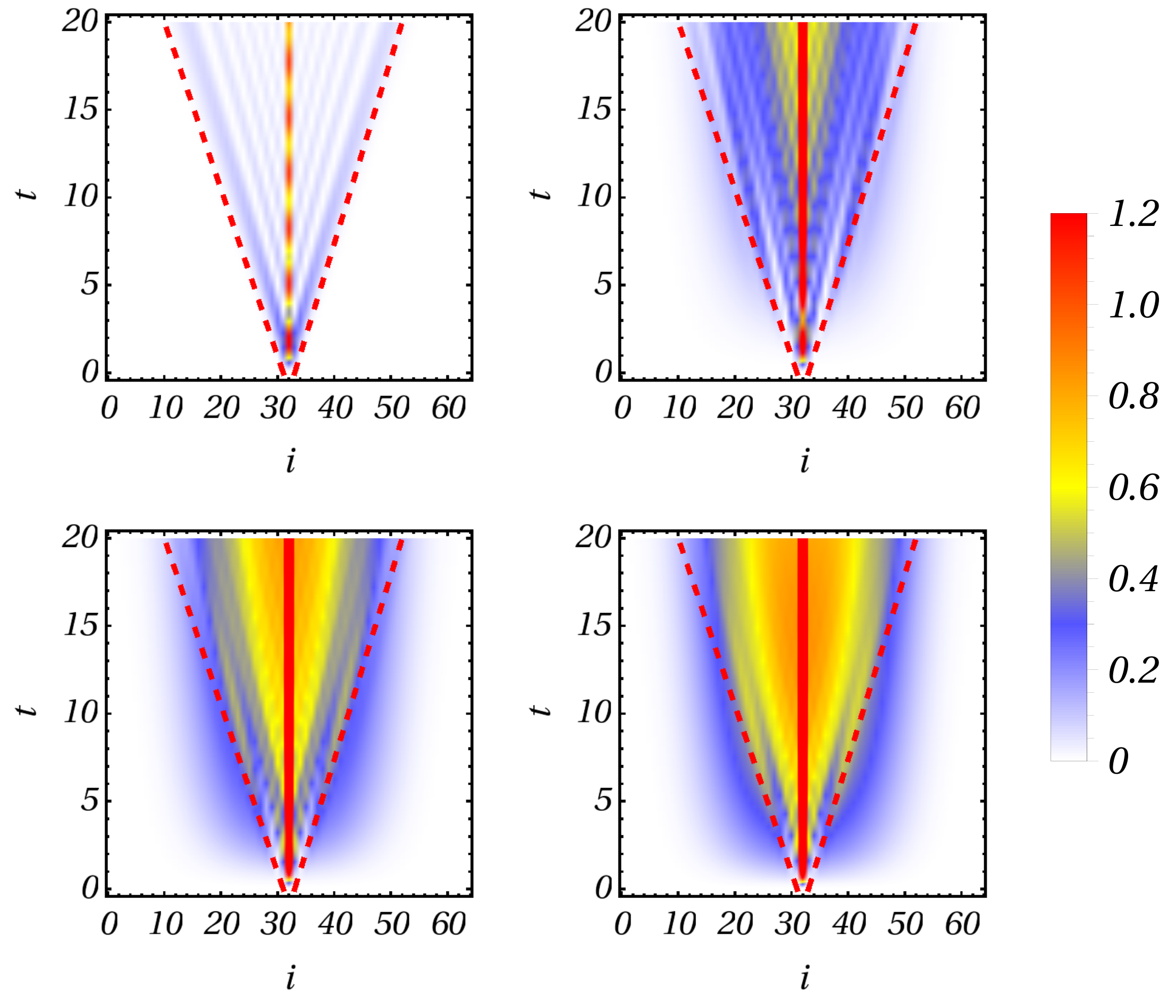}
\caption{(Color online). Density plot of the variation $|\delta G_{ii_0}^{>}(t;0)|^2$
of the particle propagator (see text) for a system with $N=63$ sites 
and $n=21$ particles initially in the ground state of the (non-interacting) system.
Different plots are for different final boson-boson interactions:
from top left to bottom right $U=0,0.1,0.3,0.5$. 
Red (dashed) lines are the reference light cone for the non-interacting case ($U=0$).}
 \label{fig:1Dggt0}
\end{figure}

In Fig.\ref{fig:1Dggt0} we plot the variation of the particle propagator 
$|\delta G_{ii_0}^{>}(t;0)|^2$ between the 
site $i_0$ and all other sites as a function of time
for the same system's parameters. 
The red dashed lines show the light-cone in the non-interacting case.
It can be seen that, as $U$ is increased, the particle propagator
acquires a diffusive behavior and that interference fringes appear
at small $U$ (upper row) due to the coupling of different modes 
by the interactions; these fringes disappear at higher interactions
for long times (bottom row) due to the interaction-induced decoherence in the system.

\subsection{Finite temperature}
It is interesting to understand the effect of the initial temperature
on the post-quench dynamics of the system, particularly because the study 
of many-body systems starting from an initial Gibbs state has attracted recently a lot
of interest in the context of quantum thermodynamics.
To study the effect of temperature on
the post-quench dynamics of the system 
we consider an initial Gibbs state of the form 
$\op{\rho}{}{}=e^{-\beta (\op{H}{0}{}-\mu \op{N}{}{})}/Tr(e^{-\beta (\op{H}{0}{}-\mu \op{N}{}{})})$
where $\op{N}{}{}=\sum\limits_i \op{b}{i}{\dag}\op{b}{i}{}$ is the total-number operator.

\begin{figure}[h]
\includegraphics[width=8cm]{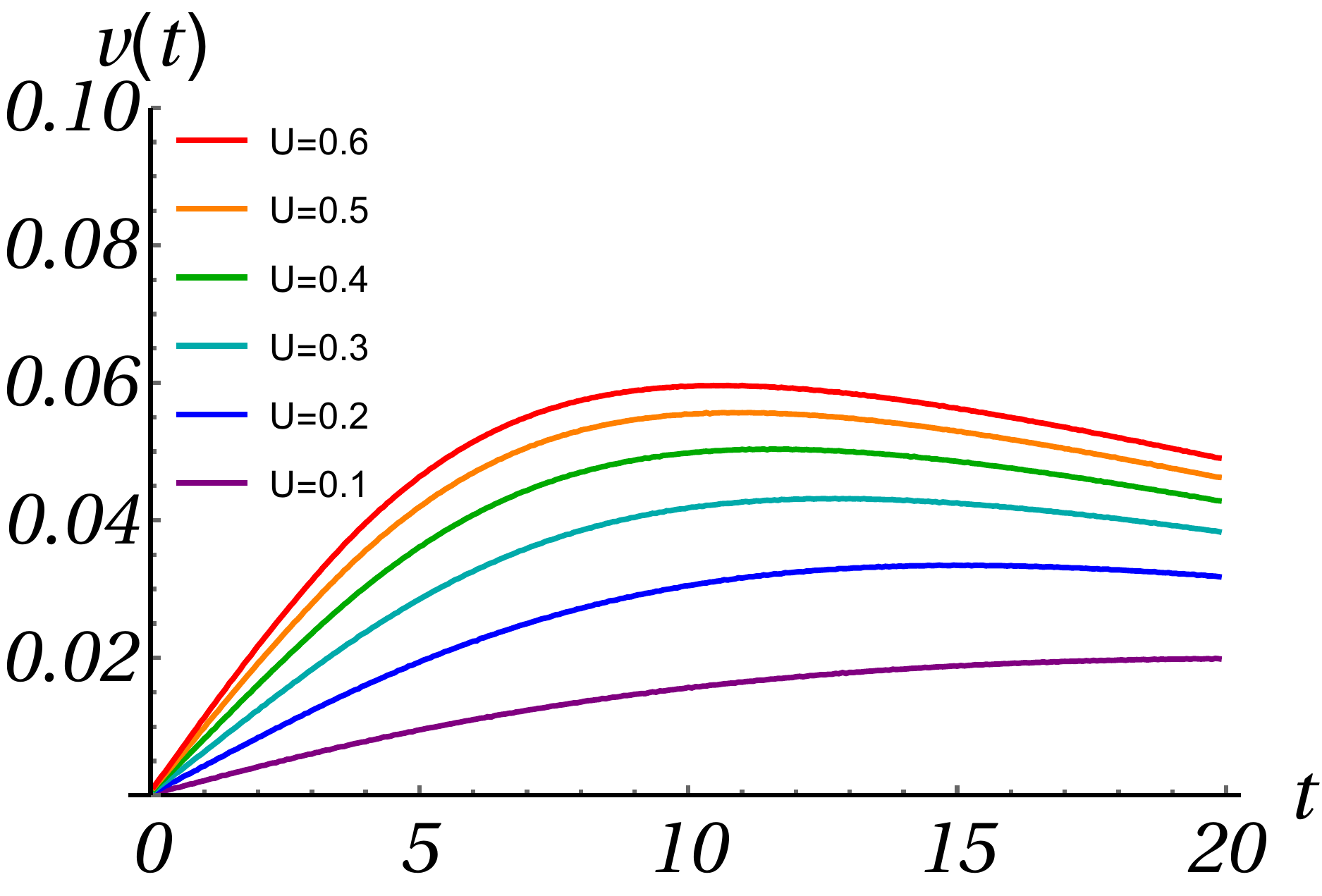}
\caption{(Color online). Velocity $v(t)$ (see text) for a system with $N=63$ sites,
$n=21$ particles. The initial state is assumed to be prepared with an initial inverse temperature
$\beta=1$. Different curves are for different final interaction strengths.}
 \label{fig:1Dvtbeta}
\end{figure}

\begin{figure}[h]
\includegraphics[width=8cm]{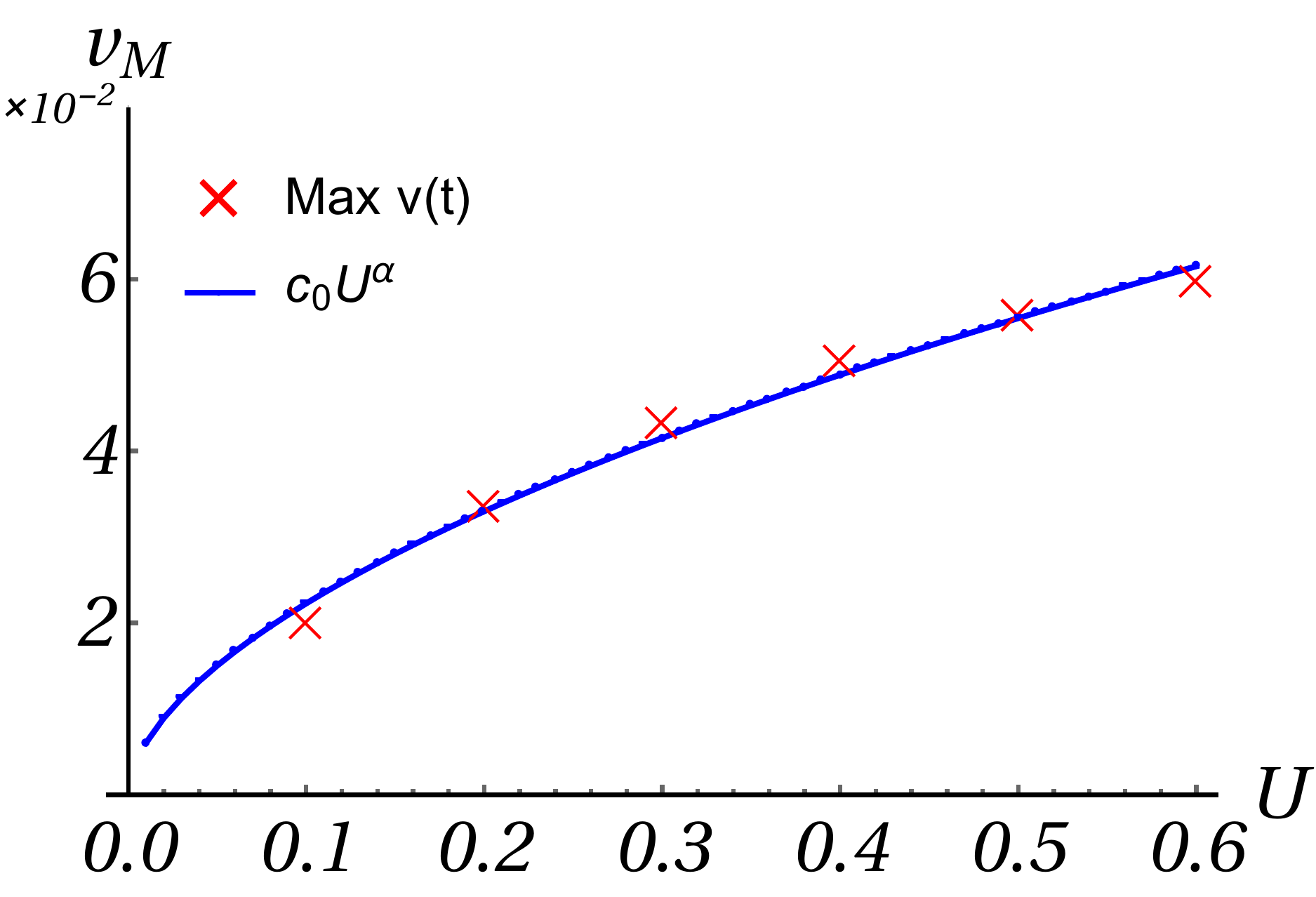}
\caption{(Color online). Maximum velocity as a function of $U$ obtained from the 
fit of the curves in Fig.\ref{fig:1Dvtbeta} with a polynomial of sixth order. 
The fitting curve is obtained as $v_{M}(U)=c_0 U^\alpha$ with $c_0=(0.082\pm 0.003)$
and $\alpha=(0.566\pm 0.038)$.}
 \label{fig:1Dvmaxbeta}
\end{figure}

In Fig.\ref{fig:1Dvtbeta} we plot the velocity of expansion of the density $v(t)$ 
for the case $\beta=T^{-1}=10$ and for a system of $N=63$ sites and an average number of bosons
$\langle\hat N \rangle=21$. Different curves are for different final interactions $U$.
Comparing corresponding curves with the ones in Fig.\ref{fig:1Dvt} we see that the effect of temperature is 
to decrease the expansion velocity.
On the other hand the argument used to predict the behavior of the maximum expansion velocity 
with respect to the interaction strength $U$ is still valid as we can observe by fitting the maximum value
for each $U$ with a power law $v_M(U)=c_0 U^\alpha$ finding $\alpha=(0.566\pm 0.038)$ (see Fig.\ref{fig:1Dvmaxbeta}).

\begin{figure}[h]
\includegraphics[width=8cm]{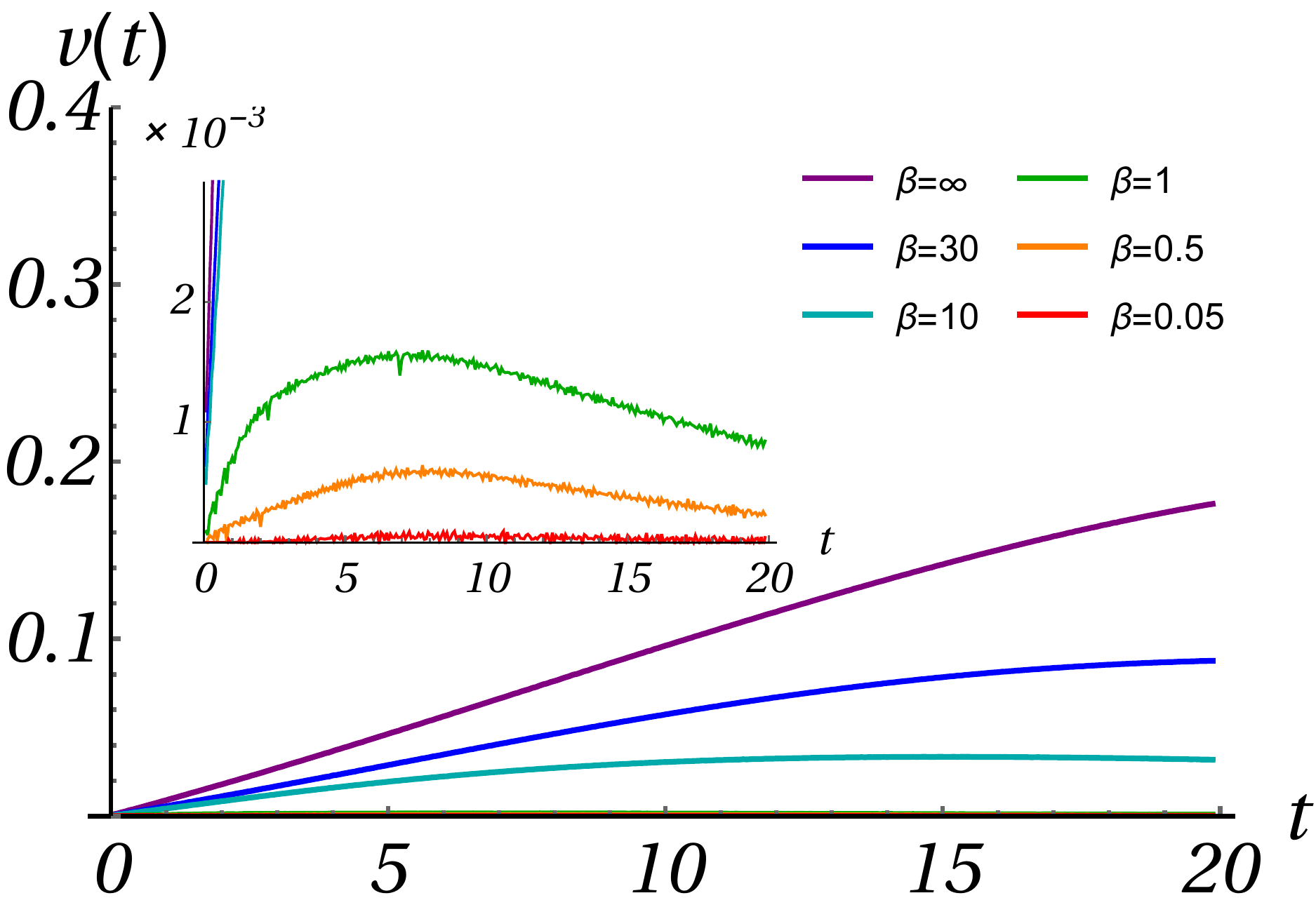}
\caption{(Color online). Velocity $v(t)$ (see text) for a system with $N=63$ sites,
$n=21$ particles and final interaction $U=0.2$.
Different curves are for different initial inverse temperature $\beta=\infty,30,10,1,0.5,0.05$.}
 \label{fig:1DvtbetaU}
\end{figure}

For fixed final interaction $U$ and for different initial inverse temperatures $\beta$
we can see from Fig.\ref{fig:1DvtbetaU} that an increase in temperature results in a 
suppression of the expansion of the system.

This behavior can be understood qualitatively
by the fact that as temperature increases, particles 
in the initial state tend to occupy 
more energy eigenstates of the non-interacting Hamiltonian and 
therefore the systems becomes effectively more dilute
in the energy eigenstates.
Therefore as the interaction is switched on the number of particles 
which are coupled decreases with the initial temperature. 
From a formal point of view the interaction Hamiltonian in the momentum 
basis (assuming periodic boundary conditions) reads: 
$\opt{V}{t}{}{}=\frac{U(t)}{2L}\;\sum_{k,p,q}\op{\tilde b}{p+q}{\dag}\op{\tilde b}{k-q}{\dag}\op{\tilde b}{p}{}\op{\tilde b}{k}{}=\frac{U(t)}{2L}\;\sum_{k,p}\op{\tilde b}{p}{\dag}\op{\tilde b}{k}{\dag}\op{\tilde b}{p}{}\op{\tilde b}{k}{}+\frac{U(t)}{2L}\sum_{k,p,q\neq 0}\cdots$.
In a dilute system as the one considered here the leading contribution is the first one (q=0) which 
contains the product $\op{\tilde n}{p}{}\op{\tilde n}{k}{}$ of the number operators for the states with momenta $p,k$.

\begin{figure}[h]
\includegraphics[width=8cm]{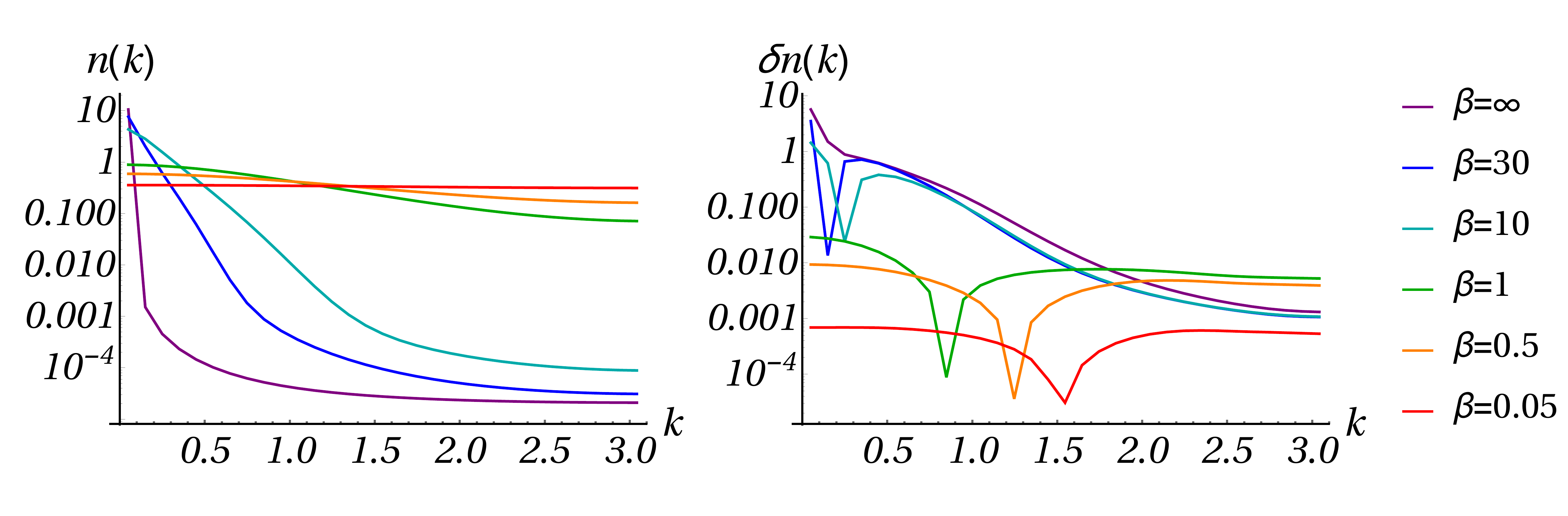}
\caption{(Color online). Initial momentum distribution $n(k)$ (left) 
and variation of the momentum distribution $\delta n(k)$ at t=20 
for a system with $N=63$ sites, $n=21$ particles and final interaction strength $U=0.2$.
Different curves are for different initial temperature $T=\beta^{-1}$ of the initial state.}
 \label{fig:1Dnkbeta}
\end{figure}

To be more quantitative let us look at the distribution of 
particles in momentum basis both at $t=0$ (initial state)
and at the end of the evolution.
Bearing this in mind we see from Fig.\ref{fig:1Dnkbeta} (left)
that at low temperatures the momentum distribution is peaked around
$k=0$ and, as expected, most of particles occupy low energy states.
On the other hand at high initial temperatures the initial momentum 
distribution is spread over the whole $k-$space
meaning that the system tends to occupy more energy levels
as it is natural to expect when temperature is increased.
We can see the effect of switching-on the interaction 
on the momentum distribution in the right panel of 
Fig.\ref{fig:1Dnkbeta}.
It shows that for low-temperature initial states
the variation of the number of particles 
at $k\approx0$ is nearly three order of magnitude
greater than for high-temperature initial states.
This means that for the latter case the system is,
within a good approximation, basically non-interacting
(as also expected by the expression for the interaction
Hamiltonian in the momentum basis).

\begin{figure}[h]
\includegraphics[width=8cm]{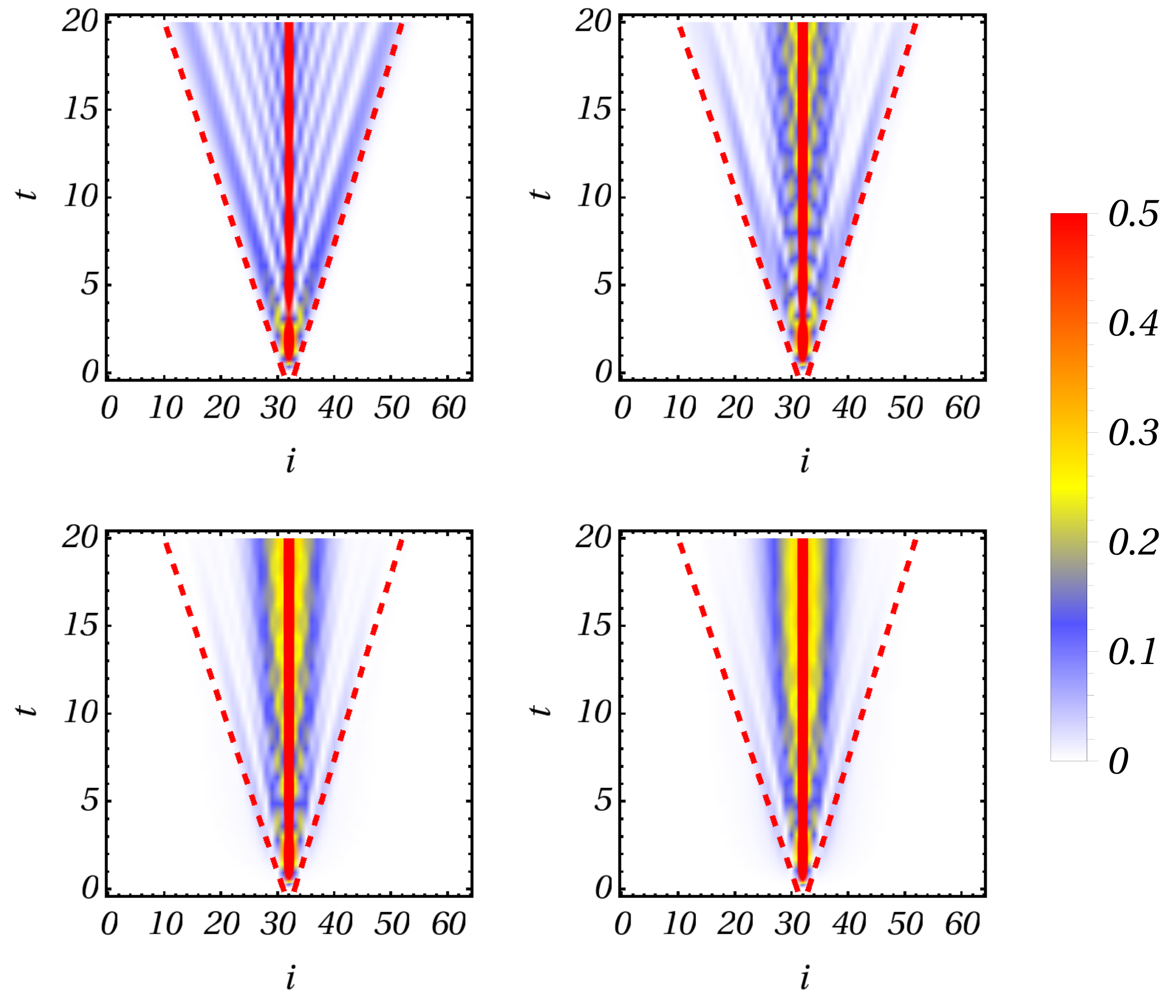}
\caption{(Color online). Density plot of the variation of the particle propagator
(see text) for a system with $N=63$ sites 
and $n=21$ particles. The initial state is a Gibbs
state of the form $\op{\rho}{}{}=e^{-\beta (\op{H}{0}{}-\mu \op{N}{}{})}/Tr(e^{-\beta (\op{H}{0}{}-\mu \op{N}{}{})})$
with inverse temperature $\beta=1$.
The chemical potential $\mu$ is chosen such that the number of particles in the 
system is $\mean{\op{N}{}{}}{\op{\rho}{}{}}=21$.
Different plots refer to different interactions, from top left to bottom 
right: $U=0,0.1,0.3,0.5$.}
\label{fig:1Dggt0beta}
\end{figure}

The propagation of correlations is also influenced by the temperature 
of the initial state.
In Fig.\ref{fig:1Dggt0beta} we show the variation of the particle propagator $\delta G^{>}(t,t^+)$
for a system with initial inverse temperature $\beta=10$ for different final interaction strengths $U=0,0.1,0.3,0.5$.
Comparing the case $U=0$ with the corresponding case at zero temperature
(Fig.\ref{fig:1Dggt0} top-left plot) we see that the maximum speed of propagation (given by the slope of the dashed red lines)
is the same. This is obviously to be expected because it only depends upon the spectrum which is the same for both
cases. By increasing the final interaction strength $U$ we again have a focalization effect and the spreading of correlations
slows down as in the case at zero temperature.

\begin{figure}[h]
\includegraphics[width=8cm]{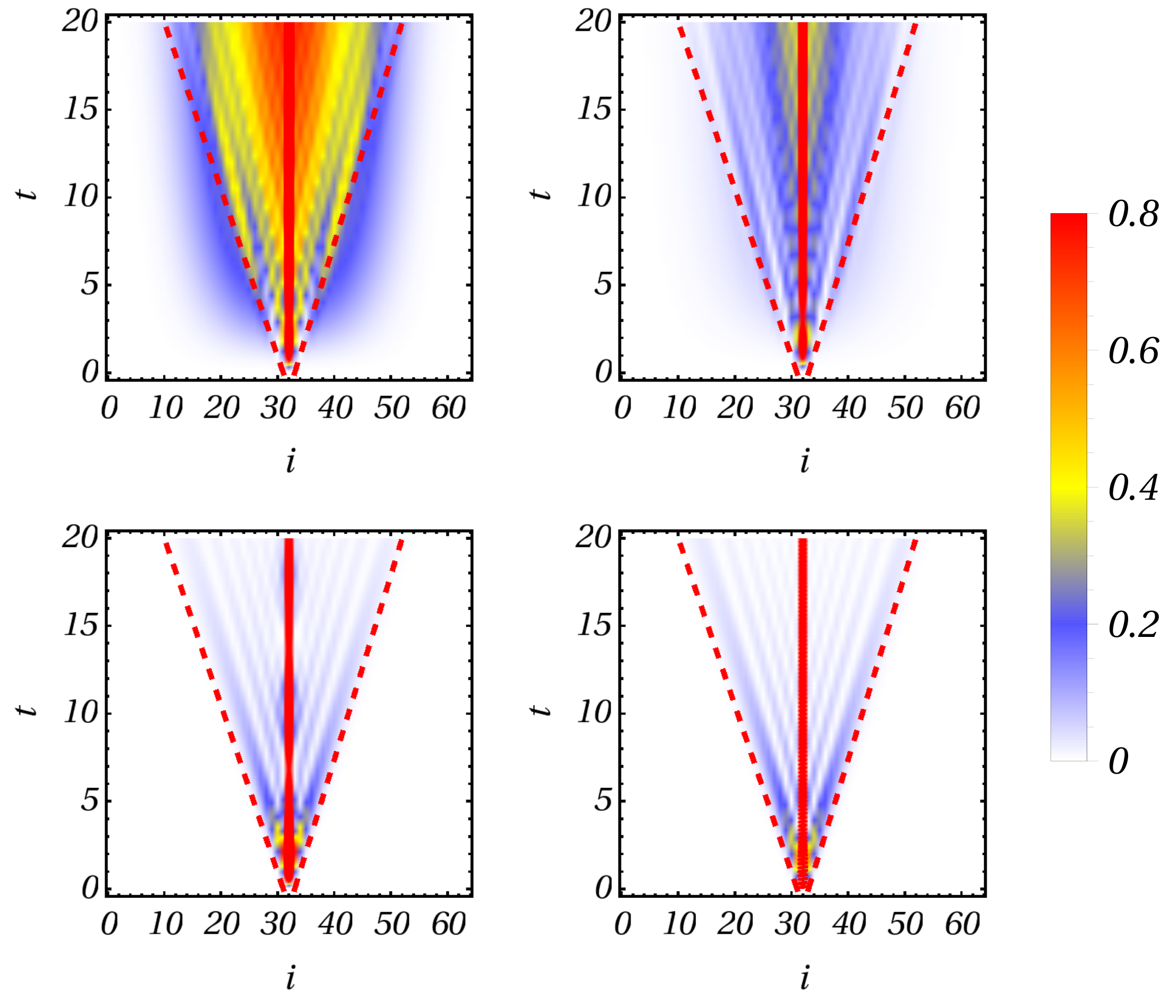}
\caption{(Color online). Density plot of the variation of the particle propagator
(see text) for a system with $N=63$ sites 
and $n=21$ particles and final interaction $U=0.2$. The initial state is a Gibbs
state of the form $\op{\rho}{}{}=e^{-\beta (\op{H}{0}{}-\mu \op{N}{}{})}/Tr(e^{-\beta (\op{H}{0}{}-\mu \op{N}{}{})})$
with inverse temperature $\beta=1$.
The chemical potential $\mu$ is chosen such that the number of particles in the 
system is $\mean{\op{N}{}{}}{\op{\rho}{}{}}=21$.
The different plots refer to different initial inverse temperature, from top left to bottom 
right: $\beta=\infty,30,1,0.05$.}
\label{fig:1Dggt0betaU}
\end{figure}

On the other hand if we fix the final interaction strength
and we look at the particle propagator for different initial temperatures
we see that by increasing it the propagation becomes more similar to the non-interacting case.
This is clearly shown in Fig.\ref{fig:1Dggt0betaU} and by comparison with the top left figure in
Fig. \ref{fig:1Dggt0}.
Therefore we can say that as the initial temperature increases the propagation of correlations
becomes ballistic.
This behavior can be explained once again by looking at the distribution of particles in the
momentum basis: at higher temperatures the effect of interactions is negligible due to the low occupancy
of each mode.

\section{Quenches in 2D Bose-Hubbard model}
\label{sec:2D}
We now move the study of the post quench dynamics of a two-dimensional
BHM on a square lattice of size $N_x\times N_y=21\times 21$.
We will see that qualitatively the results are similar to the 
one-dimensional case studied in the previous section.

\subsection{Zero Temperature}
It has been shown \cite{ronzheimer2013} that in the case of a two-dimensional 
gas of bosons on a lattice, the asymptotic velocity decreases 
by increasing the final boson-boson interaction 
for quenches from the Mott insulator phase into the superfluid one.

\begin{figure}[h]
\includegraphics[width=8cm]{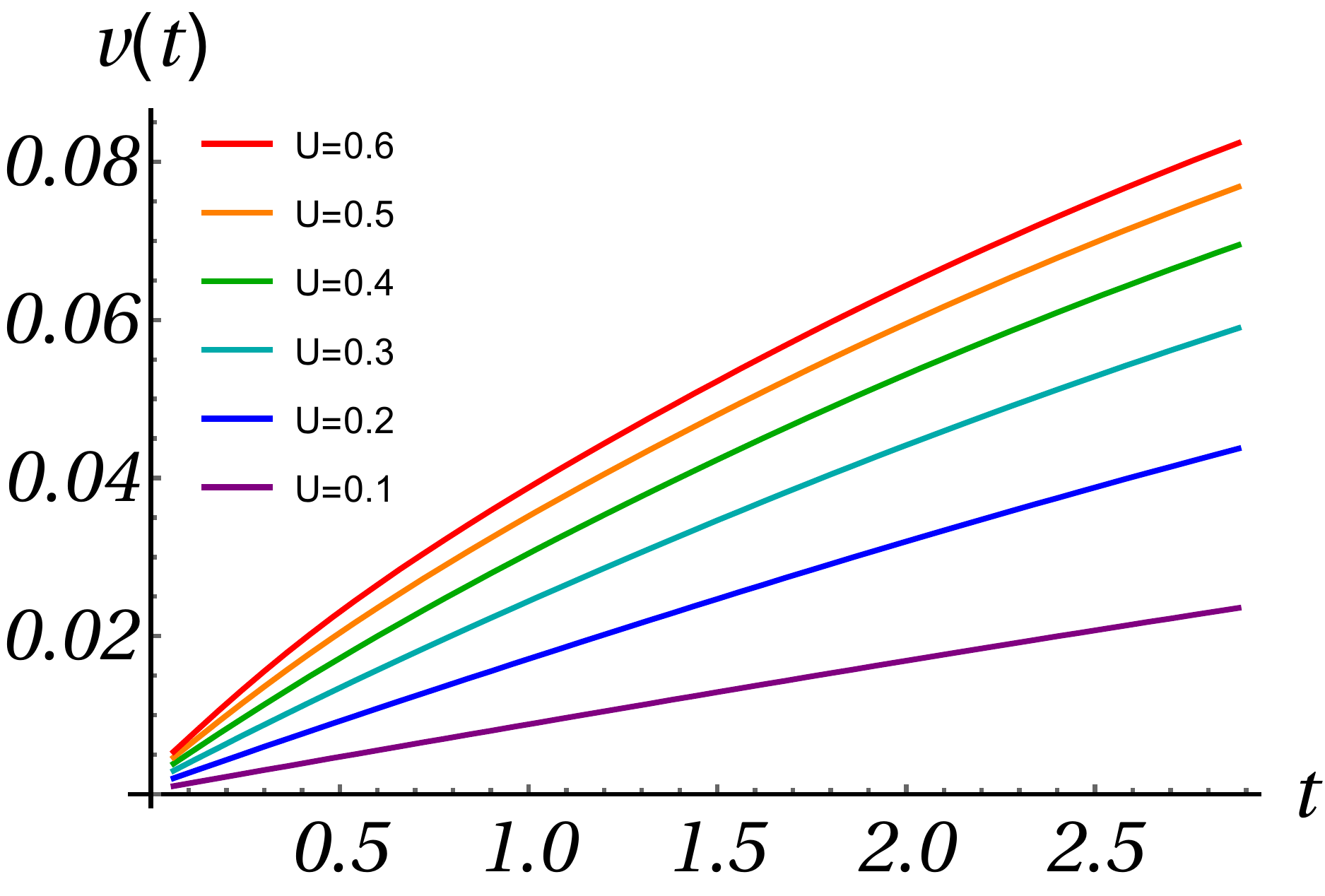}
\caption{(Color online). Velocity $v(t)$ (see text) for a two-dimensional 
the isotopic system $(J_x=J_y)$ with $N_x\times N_y=21\times 21$ sites and 
$n=147$ particles initially in the ground state of the (non-interacting) system.
Different plots are for different final boson-boson interactions.} 
\label{fig:2Dvt}
\end{figure}

\begin{figure}[h]
\includegraphics[width=8cm]{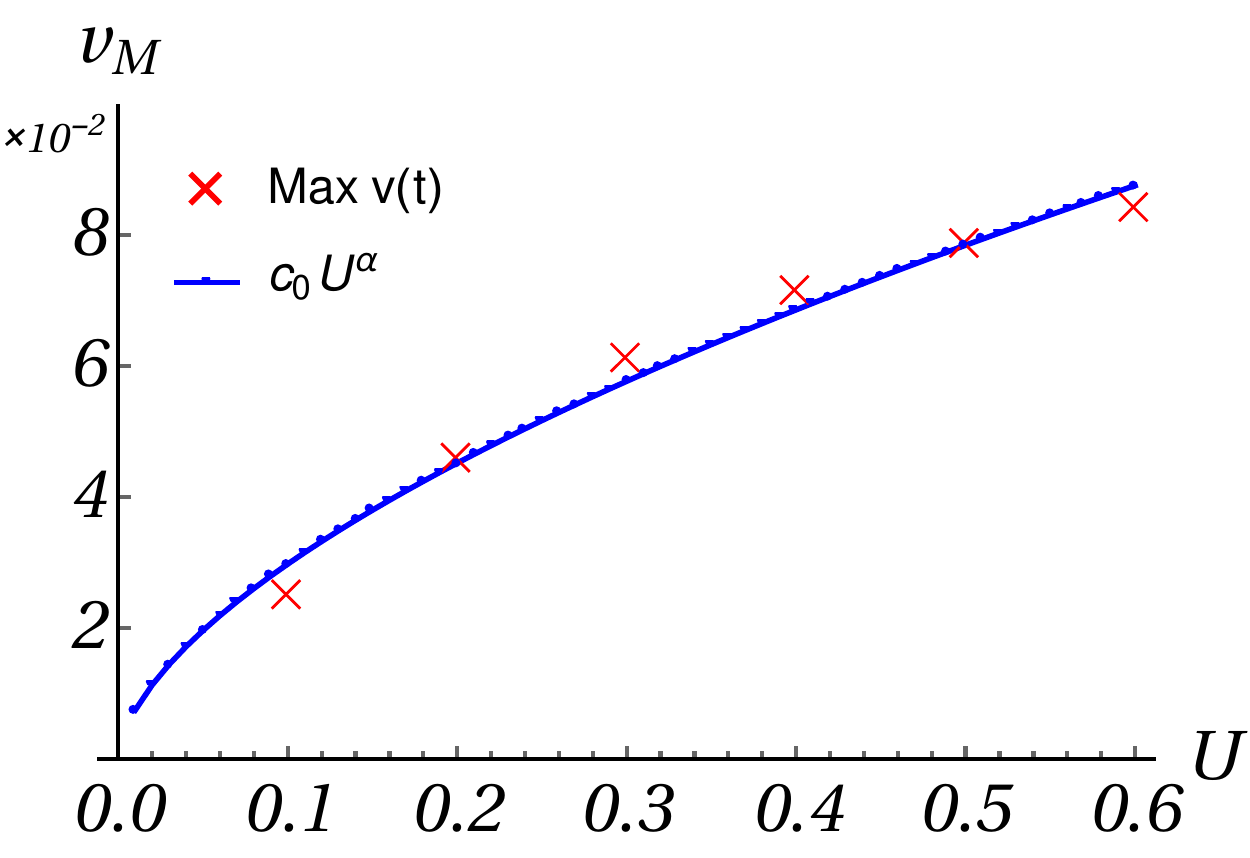}
\caption{(Color online). Maximum velocity as a function of $U$ obtained from 
the 
fit of the curves in Fig.\ref{fig:2Dvt} with a polynomial of sixth order in 
$t$. 
The fitting curve is obtained as $v_{M}(U)=c_0 U^\alpha$ with $c_0=(0.119\pm 
0.006)$ and $\alpha=(0.602\pm 0.055)$.} 
\label{fig:2Dvmax}
\end{figure}

In our case we see from Fig.\ref{fig:2Dvt} that by increasing the final 
boson-boson interaction strength the velocity increases during the transient 
as it is expected from simple energy conservation arguments.
In Fig. \ref{fig:2Dvt} we plot the velocity related to the $\sigma(t)$
relative to the motion along the $x-$axis only because the system is isotropic
and therefore the expansion is the same along any direction.
Furthermore, as in the one-dimensional case, we can see from Fig. \ref{fig:2Dvmax} 
that the maximum velocity reached has a square root dependence upon the final interaction strength which 
again confirms the conversion of the initial interaction energy injected into 
the system by the quench into kinetic energy.

\begin{figure}[h]
\includegraphics[width=8cm]{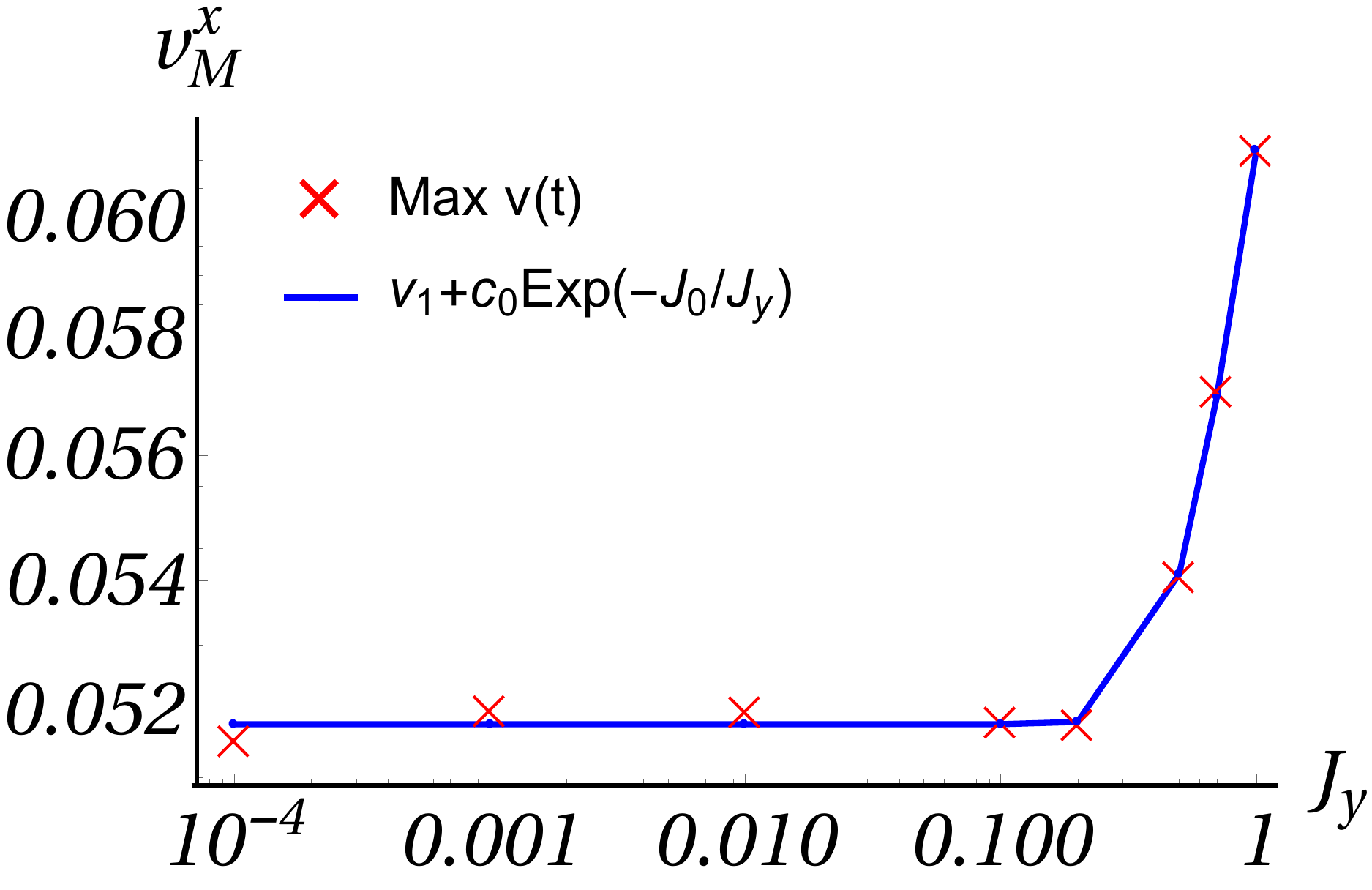}
\caption{(Color online).  Maximum velocity along the $x-$axis as a function of $J_y$ for a two-dimensional 
anisotopic system with $N_x\times N_y=21\times 21$ sites and 
$n=147$ particles initially in the ground state of the (non-interacting) system.
The fitting curve is obtained as $v_{M}(J_y)=v_1+c_0 \exp(-J_0/J_y)$ with $v_1=(0.05179\pm 
0.00007)$, $c_0=(0.038\pm 0.002)$ and $J_0=(1.400\pm 0.065)$.}
\label{fig:2DvxMjy}
\end{figure}

\begin{figure}[h]
\includegraphics[width=8cm]{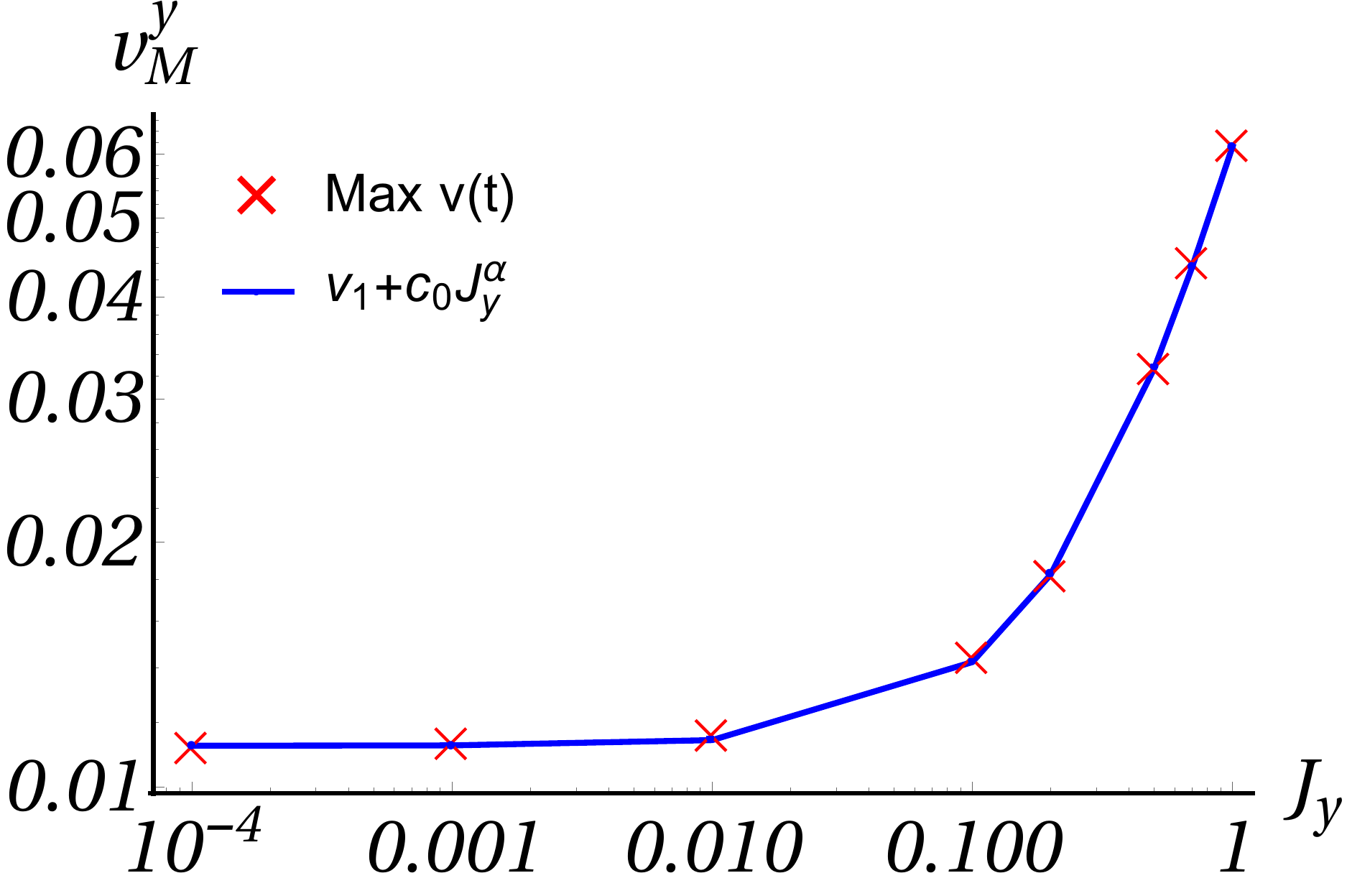}
\caption{(Color online).  Maximum velocity along the $y-$axis as a function of $J_y$ for a two-dimensional 
anisotopic system with $N_x\times N_y=21\times 21$ sites and 
$n=147$ particles initially in the ground state of the (non-interacting) system.
The fitting curve is obtained as $v_{M}(J_y)=v_1+c_0 J_y^\alpha$ with $v_1=(0.0112\pm 
0.0001)$, $c_0=(0.0500\pm 0.0002)$ and $\alpha=(1.22\pm 0.01)$.}
\label{fig:2DvyMjy}
\end{figure}

On the other hand in an anisotopic system ($J_y\neq J_x$) 
the velocity is expected to be different along different directions.
In Fig.\ref{fig:2DvxMjy} and Fig.\ref{fig:2DvyMjy} we see that the maximum velocity
along the $x-$ and $y-$axis respectively is different not only in the value,
but also on its dependence upon the tunneling rate along the $y-$axis.
In particular by tightening the trap along the $y-$axis the maximum velocity along the $x-$direction
is suppressed exponentially with $1/J_y$ whereas the maximum velocity along the $y-$axis has a power law 
dependence upon $J_y$.

\begin{figure}
\includegraphics[width=8cm]{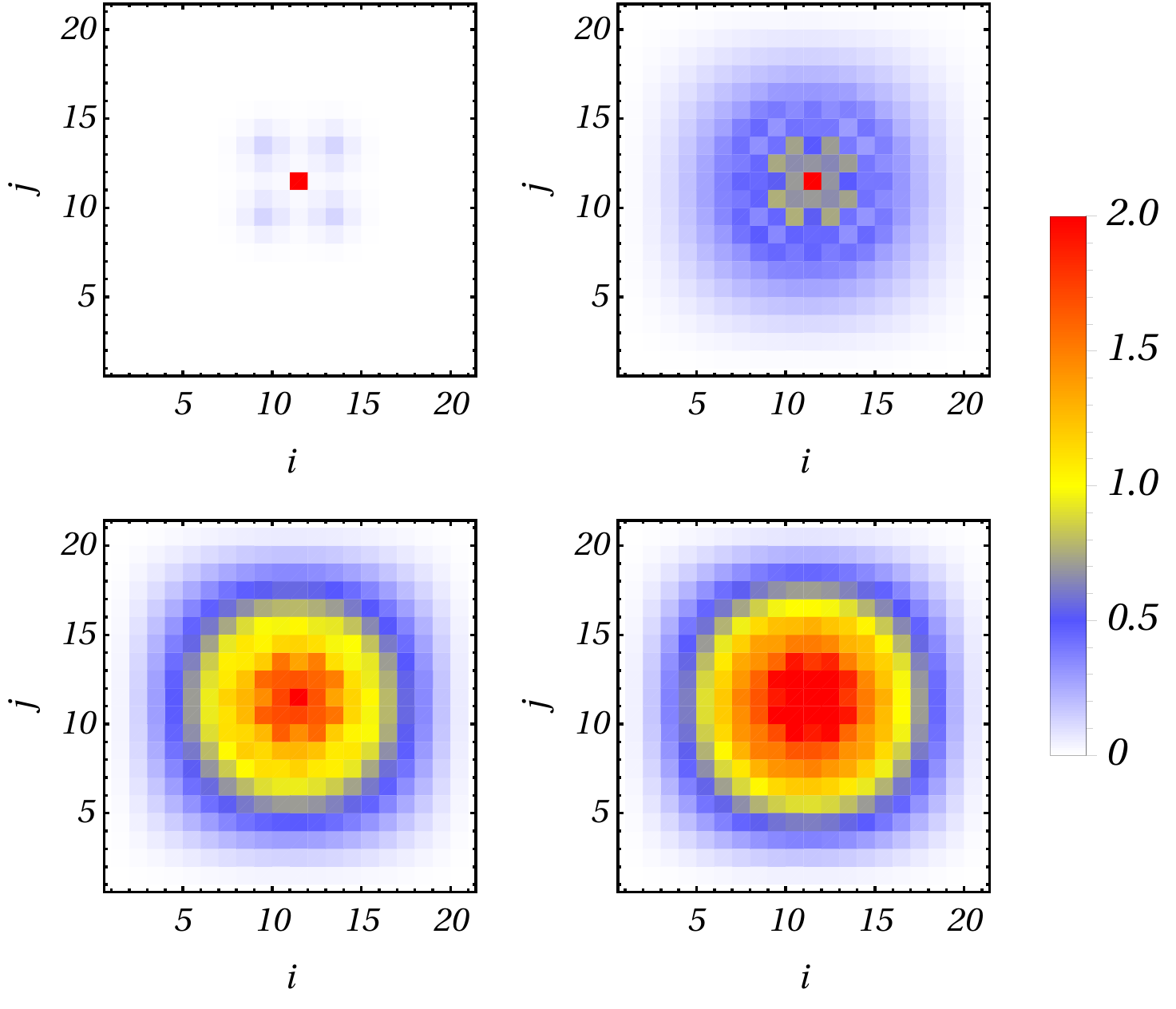}
\caption{(Color online). 
Density plot of the variation $|\delta G_{ii_0}^{>}(t;0)|^2$
of the particle propagator (see text) for a two-dimensional 
the isotopic system $(J_x=J_y)$ with $N_x\times N_y=21\times 21$ sites and 
$n=147$ particles initially in the ground state of the (non-interacting) 
system. Different plots are for different final boson-boson interactions:
from top left to bottom right $U=0,0.1,0.3,0.5$ at $t=3$.}
 \label{fig:2Dggt0vsU}
\end{figure}

It is also interesting to study the r\^ole of dimensionality
in the propagation of correlations, namely the crossover from
an isotropic two dimensional lattice $(J_x=J_y)$
to an anisotropic one $(J_x>J_y)$.
In the one-dimensional case we found that the propagation of correlations
turns from ballistic to diffusive as the final interaction strength increases.
In two dimensions we have a similar behavior as seen in Fig. \ref{fig:2Dggt0vsU} 
where we show the variation of the particle propagator $|\delta G_{{\bf r}{\bf r}_0}^{>}(t;0)|$, 
where now ${\bf r}=(i,j)$ and ${\bf r}_0=(\lfloor N/2\rfloor+1,\lfloor N/2\rfloor+1)$, 
for an isotropic system $(J_x=J_y)$ with $N_x \times N_y =21 \times 21$ sites.
All plots refer to $t=3$ whereas from top left to bottom right the interaction strength 
takes the values $U=0.005,0.1,0.2,0.4$.
We can see that at small interactions (top row) a square ''wave front''
spreads ballistically from the center whereas at higher interactions (bottom row)
the propagation becomes diffusive as witnessed by a guassian-like distribution.

\begin{figure}
\includegraphics[width=8cm]{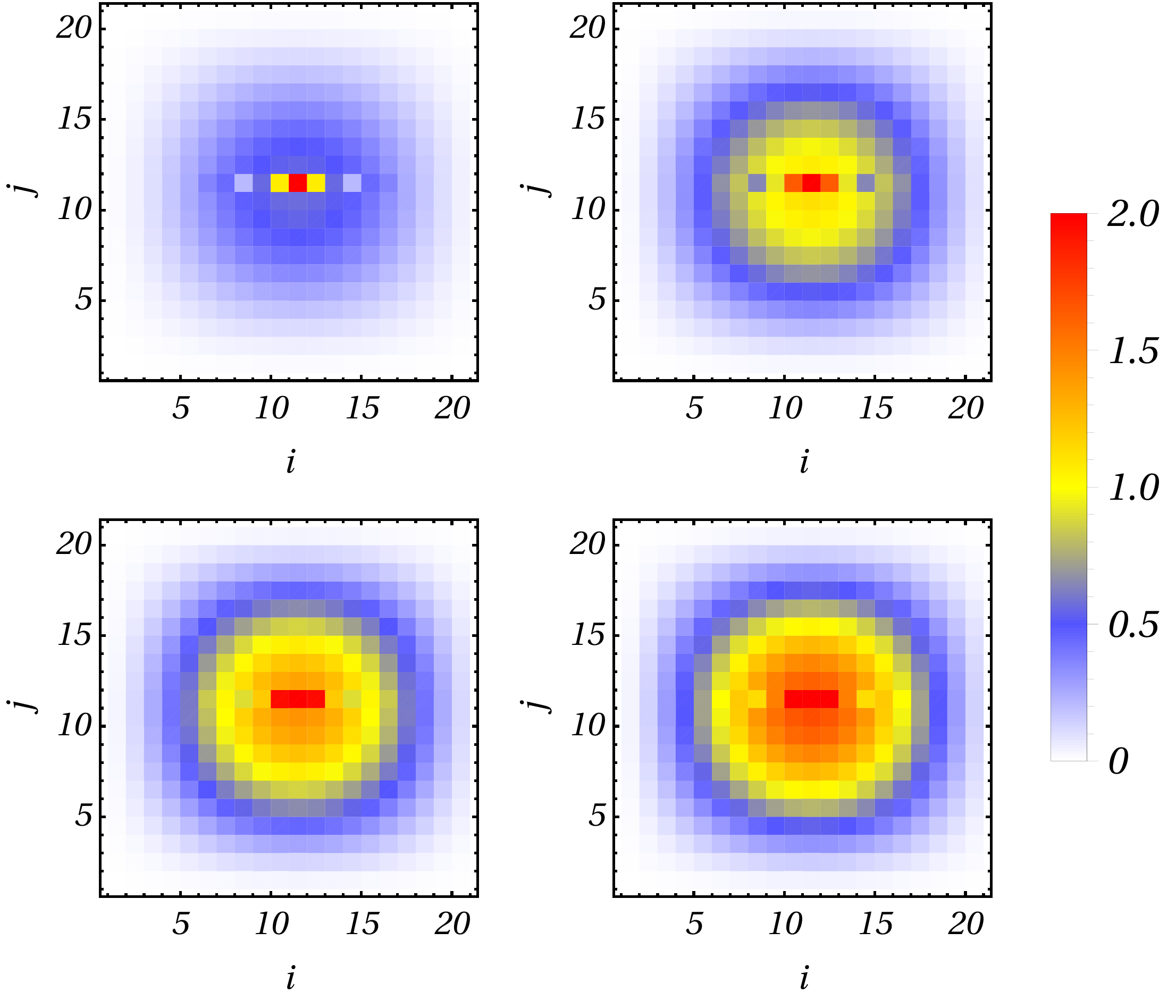}
\caption{(Color online). Density plot of the variation
of the particle propagator (see text) for an anisotropic two-dimensional
system with $N_x\times N_y=21\times 21$ sites 
and $n=147$ particles initially in the ground state of the (non-interacting) system.
The tunneling rate along the $y$-direction in units of the tunneling rate along the $x$-direction $J_x$ is $J_y=0.001$ 
Different density plots correspond to different final interactions (from top left to bottom right) $U=0.1,0.2,0.3,0.4$
at $t=3$.}
 \label{fig:2Danigg0t}
\end{figure}

In an anisotropic system the spreading of
correlations has different behaviors along the two axis 
as it would have been expected.
This can be clearly seen in Fig.~\ref{fig:2Danigg0t} 
where we considered the case $J_y/J_x=10^{-3}$.
At small interaction strengths we can clearly see the difference between the propagation
in the two directions whereas as the interaction is increased the propagation
tends to become more homogeneous.
This can be explained by the following argument:
in the limit $J_y\rightarrow 0$ the system behaves as weakly coupled 
one dimensional Bose-Hubbard models in the $y-$direction.
The gas is therefore made of one dimensional ''pipes'' with lower density
than the total one.
Since the effect of the interaction depends crucially on the density of the system
the expansion velocity decreases 
with the initial density of the system and therefore also in this case the average velocity
tends to decrease in the limit $J_y\rightarrow 0$, as shown in Fig. 
\ref{fig:2Dvt}, in agreement with recent findings \cite{fabrizio2014} obtained 
by Gutzwiller ansatz.

\subsection{Finite Temperature}

\begin{figure}[h]
\includegraphics[width=8cm]{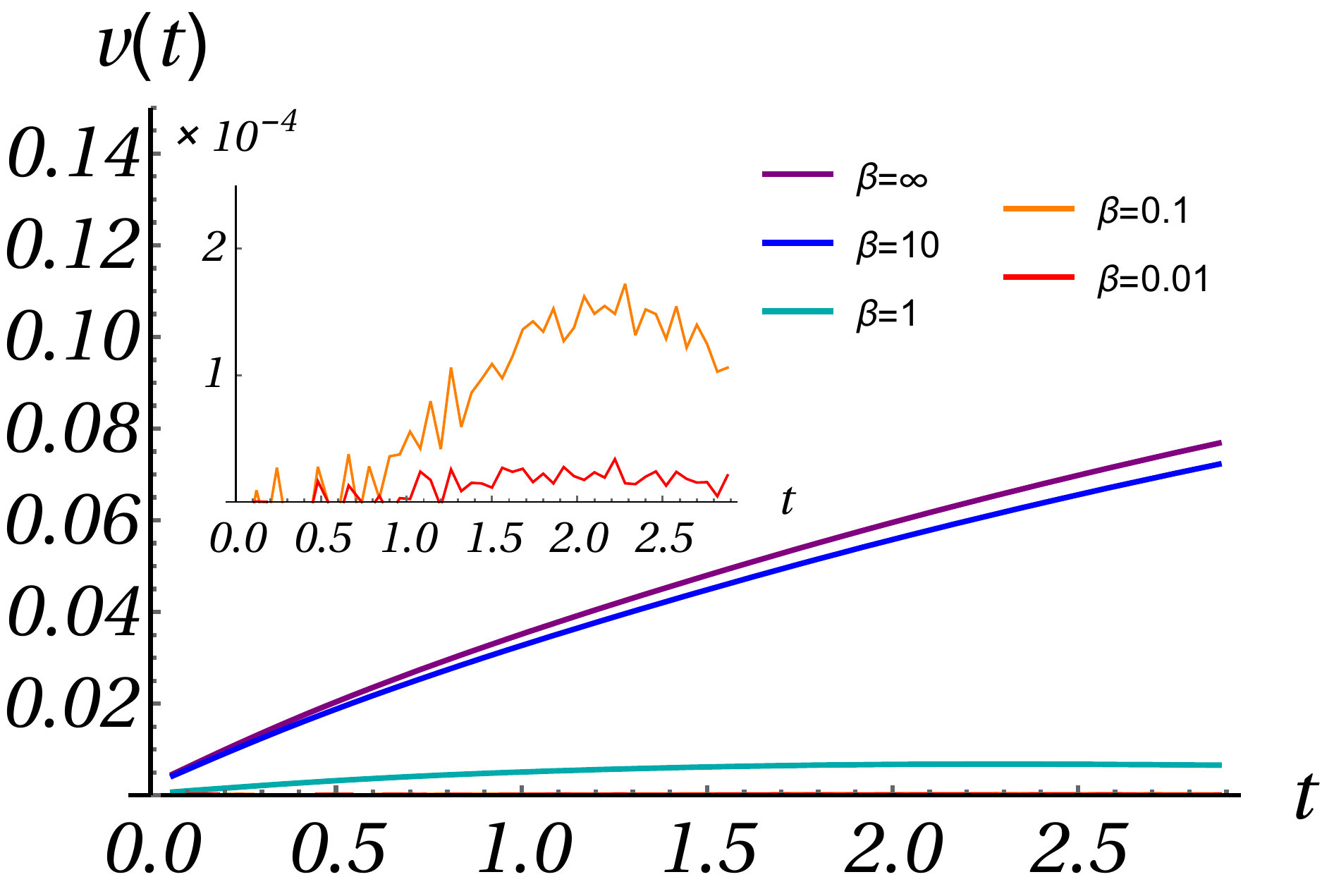}
\caption{(Color online). Velocity $v(t)$ (see text) along the x-direction for an isotrpic 
two-dimensional BHM with $N_x\times N_y=21\times 21$ sites,
$n=147$ particles. Different curves are for different initial temperatures $T=\beta^{-1}$.
In the inset we show a zoom in for the curves at high initial temperatures. The final interaction strength is
set to $U=0.5$.}
 \label{fig:2Dvtbeta}
\end{figure}

\begin{figure}[h]
\includegraphics[width=8cm]{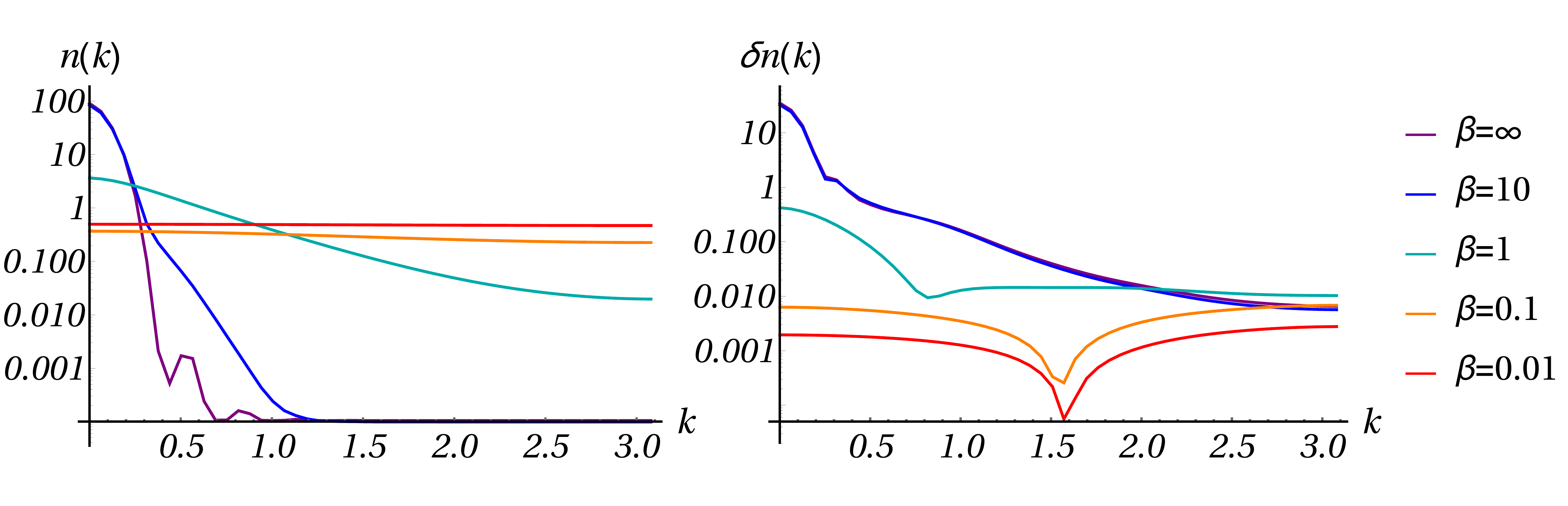}
\caption{(Color online). Initial momentum distribution $n(k)$ (left panel) 
and variation of the momentum distribution $\delta n(k)$ at t=3 (right panel)
for a system with $N_x\times N_y=21 \times 21$ sites,
$n=147$ particles and final interaction $U=0.5$.
We plot a cut of the two dimensional momentum distribution for $k_x=k_y=k$
where $k_x$ and $k_y$ are the wavevector in the $x-$ and $y-$ direction respectively.
Different curves are for different initial temperature $T=\beta^{-1}$ of the initial state.}
 \label{fig:2Dnkvsbeta}
\end{figure}

We have seen in Sec.\ref{sec:1D} that for a 1D BHM
the effect of the initial temperature is to 
slow down the expansion of the density following a quench in the interaction. 
At very high temperature the expansion is even slower than that in
the non-interacting case.
In two dimensions we have a similar effect as it can be seen from Fig.\ref{fig:2Dvtbeta}.
Similarly to the one dimensional case the reason for different behaviors at 
low and high initial temperature cases resides in the initial
occupancy of higher energy levels (left panel in Fig. \ref{fig:2Dnkvsbeta}). 
At fixed average number of particles, the temperature 
makes the system more dilute in the energy levels therefore reducing the 
scattering amplitude between different energy eigenstates
as witnessed by the difference in the final and initial momentum distribution
shown in the right panel of Fig. \ref{fig:2Dnkvsbeta}.

\begin{figure}[h]
\includegraphics[width=8cm]{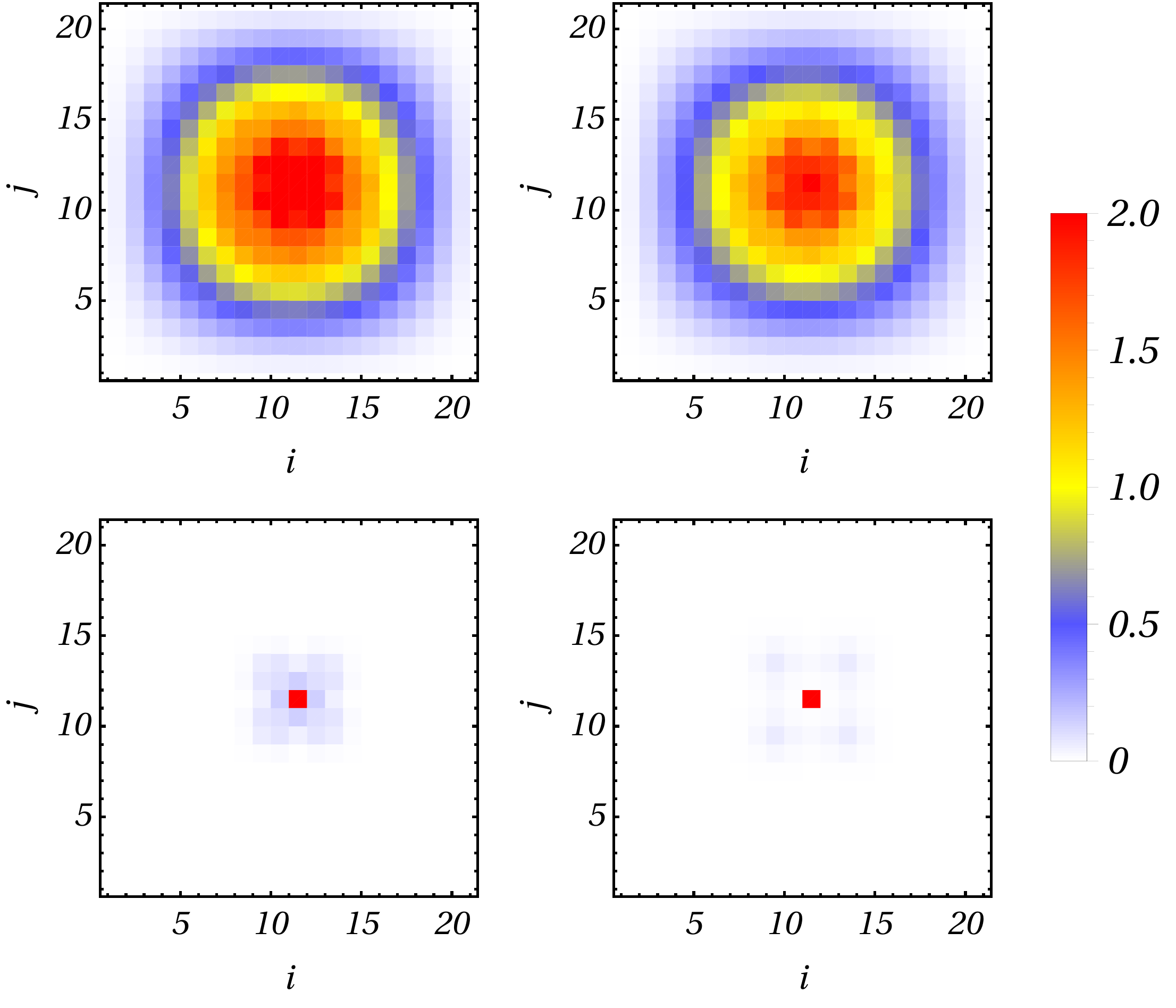}
\caption{(Color online). Density plot of the variation of the particle propagator
(see text) for a isotropic two-dimensional
system with $N_x\times N_y=21\times 21$ sites 
, $n=147$ $(n/N=1/3)$ particles, and final interaction $U=0.5$. The plots are taken at time $t=3.0$.  
The initial state is a Gibbs
state of the form $\op{\rho}{}{}=e^{-\beta (\op{H}{0}{}-\mu \op{N}{}{})}/Tr(e^{-\beta (\op{H}{0}{}-\mu \op{N}{}{})})$
with inverse temperature (from top left to bottom right) $\beta=\infty,10,1,0.1$. 
The chemical potential $\mu$ is chosen such that the number of particles in the system is $\mean{\op{N}{}{}}{\op{\rho}{}{}}=147$.}
 \label{fig:2Dgg0tbeta}
\end{figure}

In two dimensions the temperature plays a similar r\^ole
as one can see in Fig. \ref{fig:2Dgg0tbeta} where 
it is shown that by increasing the temperature the propagation of correlations
turns from diffusive (top row) to ballistic (bottom row). 

\section{Conclusions}
We have presented an approach for studying the dynamics of weakly interacting Bose-Hubbard model based on the iterative solution of a set of equations for the single 
and two particle Green's functions in the framework of the non-equilibrium Green's functions (Schwinger-Keldysh formalism).
The advantage of such an approach with respect to other widely used techniques such as
exact numerical diagonalization and 
time dependent density matrix renormalization group
is that it allows for the calculation of two-time correlation functions for 
relatively large systems in one and two dimensions. 
We applied this approach to study global quenches in the 
interaction parameter for one- and two-dimensional BHM,  
finding the crossover from ballistic to diffusive regime in the propagation of 
correlations and the temperature effects in the dynamics. 
The approach presented here 
is also suitable for extensions in different directions
such as the study of the dynamics from a general interacting initial state by
allowing the contour $\gamma$ to have a branch over the imaginary axis 
\cite{stefanucci},
the study of strong coupling limit \cite{kennett2011,kennett2016},
the study of transport in the presence of thermal reservoirs \cite{wang2006}.

\section{Acknowledgments}
We thank Margherita Marsili for useful discussions and 
acknowledge financial support from MIUR through FIRB Project No. RBFR12NLNA\_002.
NLG acknowledges financial support from the EU collaborative project QuProcS
(Grant Agreement 641277). We thank the Department of Physics 
of the Universit\'a della Calabria for providing access to 
high performance computational resources through the NEWTON cluster
(Progetto PONa3\_00370 Materiali, Tecnologie e Ricerca Avanzata - MaTeRiA).

\bigskip
\begin{appendix}
\section{Formal solution for the real time components}
\label{app:realtime}
In this section we give the formal solution for the 
Keldysh components of the interacting single particle Green's functions.
We start from the Bethe-Salpeter equation for the kernel $K$.
We have seen that the only non-vanishing terms are 
$K_{i_1 i_1;i_2 i_2}(z_1;z_2)$ and thus we can write it
as the product $K_{i_1 j_1;i_2 j_2}(z_1;z_2)=k_{i_1;i_2}(z_1;z_2)\delta_{i_1,j_1}\delta_{i_2,j_2}$
with $k_{i_1;i_2}(z_1;z_2)=K_{i_1 i_1;i_2 i_2}(z_1;z_2)$.
We can thus rewrite the Bethe-Salpeter equation as
a Fredholm equation::

\begin{eqnarray}
 k_{i_1i_2}(z_1;z_2)&=&U(z_1)\delta_{i_1i_2}\delta_{\gamma}(z_1-z_2)\\
 &&+\sum_{\substack{\overline{i}_1}}\int_{\gamma}d \overline{z}_1\;A_{i_1;\overline{i}_1}(z_1;\overline{z}_1)\; k_{\overline{i}_1i_2}(\overline{z}_1;z_2)\nonumber,
 \label{eq:fredeq}
 \end{eqnarray}

 where we defined the kernel of the integral equation as 
 $A_{i_1i_2}(z_1;z_2)= \imath U(z_1)\; g_{i_1i_2}(z_1;z_2)\;g_{i_1i_2}(z_1;z_2)$.

Using Langreth's theorem we can write the equations
satisfied by the retarded $(R)$, advanced $(A)$, 
lesser $(<)$ and greater $(>)$ components:

\begin{eqnarray}
 \label{eq:keldcomk}
 k^{R/A}&=&U\delta+A^{R/A}\circ k^{R/A}\\
 k^{\lessgtr}&=&A^{R}\circ k^{\lessgtr}+A^{\lessgtr}\circ k^{A},
\end{eqnarray}

where we dropped the indexes and introduced the symbol $\circ$
which replaces the sums and integrals.
The above equations have solutions:

\begin{eqnarray}
 \label{eq:keldcomk}
 k^{R/A}&=&U \;(1-A^{R/A})^{-1}\\
 k^{\lessgtr}&=&(1-A^{R})^{-1}\circ A^{\lessgtr}\circ k^{A},
\end{eqnarray}
whereas for the components of the self energy we have:
\begin{widetext}
\begin{eqnarray}
 \Sigma_{i_1i_2}^{\lessgtr}(t_1;t_2)&=&2\imath \;k_{i_1i_2}^{\lessgtr}(t_1;t_2)g_{i_2i_1}^{\gtrless}(t_2;t_1)\\
 {\Sigma^{R}}_{i_1i_2}(t_1;t_2)&=&\Theta(t_1-t_2)\left({\Sigma^{>}}_{i_1i_2}(t_1;t_2)-{\Sigma^{<}}_{i_1i_2}(t_1;t_2)\right)\\
 {\Sigma^{A}}_{i_1i_2}(t_1;t_2)&=&\Theta(t_2-t_1)\left({\Sigma^{<}}_{i_1i_2}(t_1;t_2)-{\Sigma^{>}}_{i_1i_2}(t_1;t_2)\right).
\label{eq:keldcomself}
\end{eqnarray}
\end{widetext}
It is now easy to find the solution for the different components of the Dyson equation
giving the different components of the interacting single-particle Green's functions:

\begin{widetext}
\begin{eqnarray}
{G^{R/A}}&=&\left(1-{g^{R/A}}\circ{\Sigma^{R/A}}\right)^{-1}\circ {g^{R/A}}\nonumber \\
{G^{\lessgtr}}&=&\left(1-{g^{R}}\circ{\Sigma^{R}}\right)^{-1}\circ\left({g^{\lessgtr}}+{g^{R}}\circ{\Sigma^{\lessgtr}}\circ{G^{A}}+{g^{\lessgtr}}\circ{\Sigma^{A}}\circ{G^{A}}\right)\nonumber.
\end{eqnarray}
\end{widetext}

\section{Comparison with exact solution}
\label{app:exact}
Here we compare the dynamics of a one dimensional
BHM as ottained by the self-consistent ladder aproximation
with the results obtained by exact diagonalization. 
Due to the heavy computational requirements of the exact
diagonalization, we restrict the comparison 
to the zero temperature case and small sized systems. 
Furthermore we will only look at of equal time correlation functions
such as numeber of bosons per site and single particle reduced density matrix.
The exact diagonalization is done by considering $n$ bosons in $N$
sites allowing for a maximum number of bosons at each site $m_i=n$.
Therefore the exact solution does not suffer from 
reduction of the whole Hilbert space with $n$ bosons to
some of its subspaces ({\it e.g. $m_i<n$}).
The price to pay for this is
obviously that we cannot study the dynamics of too large systems 
and/or with too many bosons.

In the main text we studied the post-quench dynamics
in homogenous systems with open boundary conditions. 
Here, in order to compare the dynamics given by our
approach to the exact one we will consider both this case (homogeneous) 
and a one with a slightly richer dynamics.
Specifically we will assume the presence of an harmonic trap at the left 
boundary of the system ($\epsilon_i = 5\times10^{-3} (i-2)^2$).
In this case the ground state of the Hamiltonian shows a distribution around the minima
of the potential and, at zero temperature, all bosons initially 
occupy this state at $t=0^{-}$.
The switching on of the (repulsive) boson-boson interaction at $t=0$
will make the cloud expand.
Due to the open boundary conditions, all bosons will tend to go towards the
right end of the system ($i>2$) thus developing a non-zero total momentum
which comes from the reflection at the left boundary.

\begin{figure}
\begin{tabular}{c|c}
\includegraphics[width=4cm]{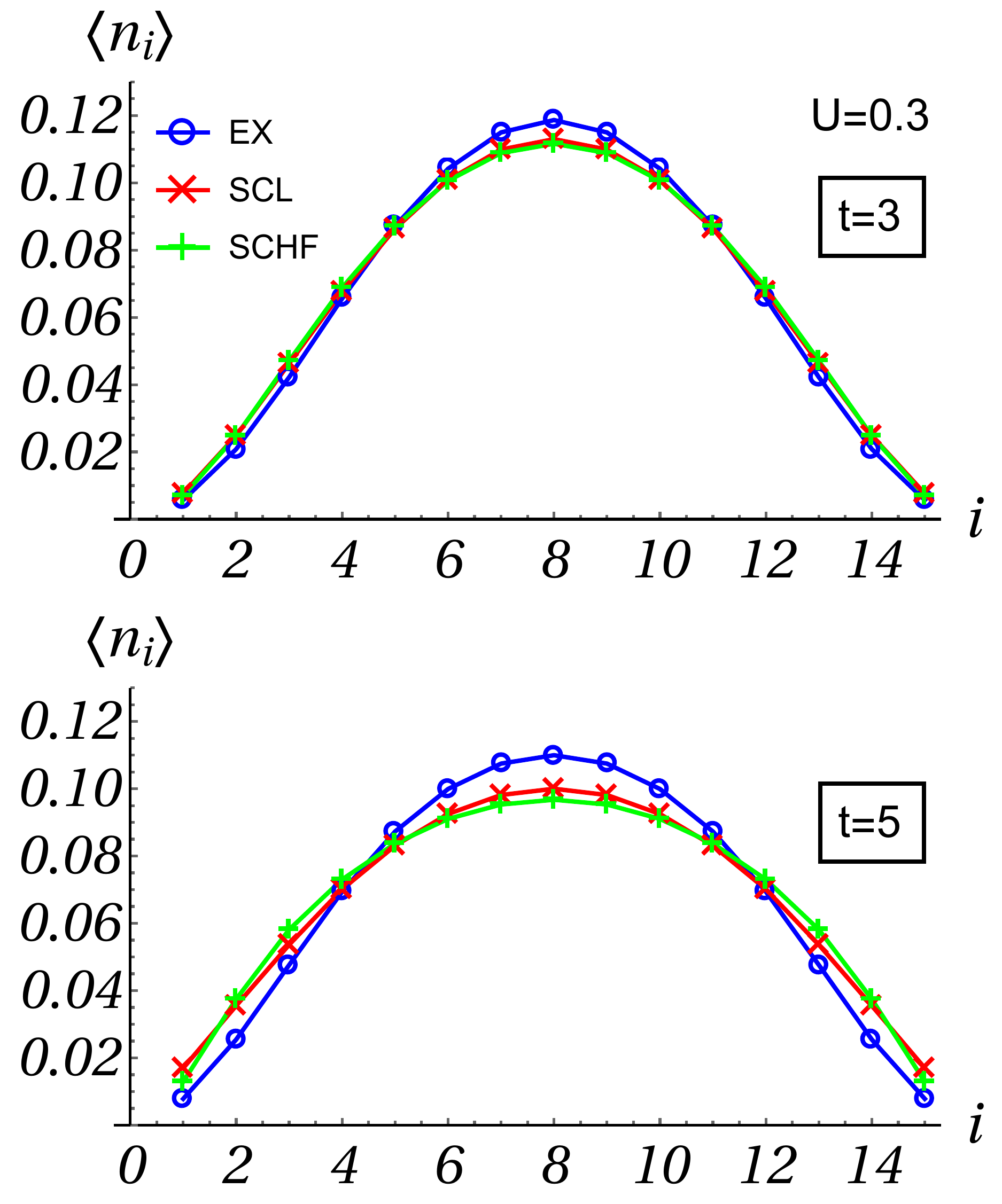}&\includegraphics[
width=4cm]{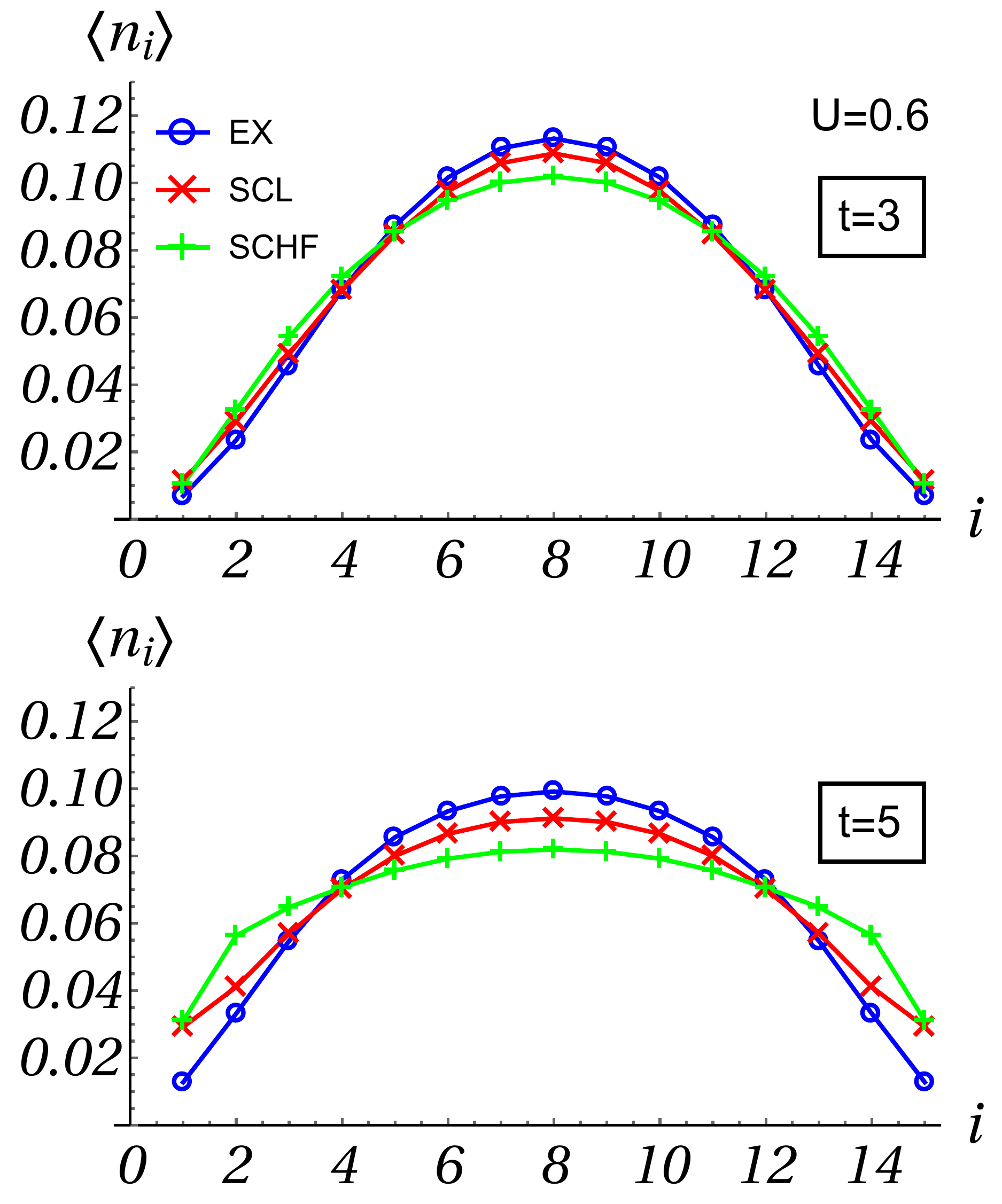}
\end{tabular}
 \caption{(Color online). Average number of bosons per site at two different 
times (upper row) $t=3$ and (lower row) $t=5$ and for two different final 
interactions (left column) $U=0.3$ and (right column) $U=0.6$. Here for a homogeneous system (see text) 
with $N=15$ sites, $n=5$ initially in its non-interacting ground state $U=0$.}
 \label{fig:numx-ns15}
\end{figure}

\begin{figure}
\begin{tabular}{c|c}
\includegraphics[width=4cm]{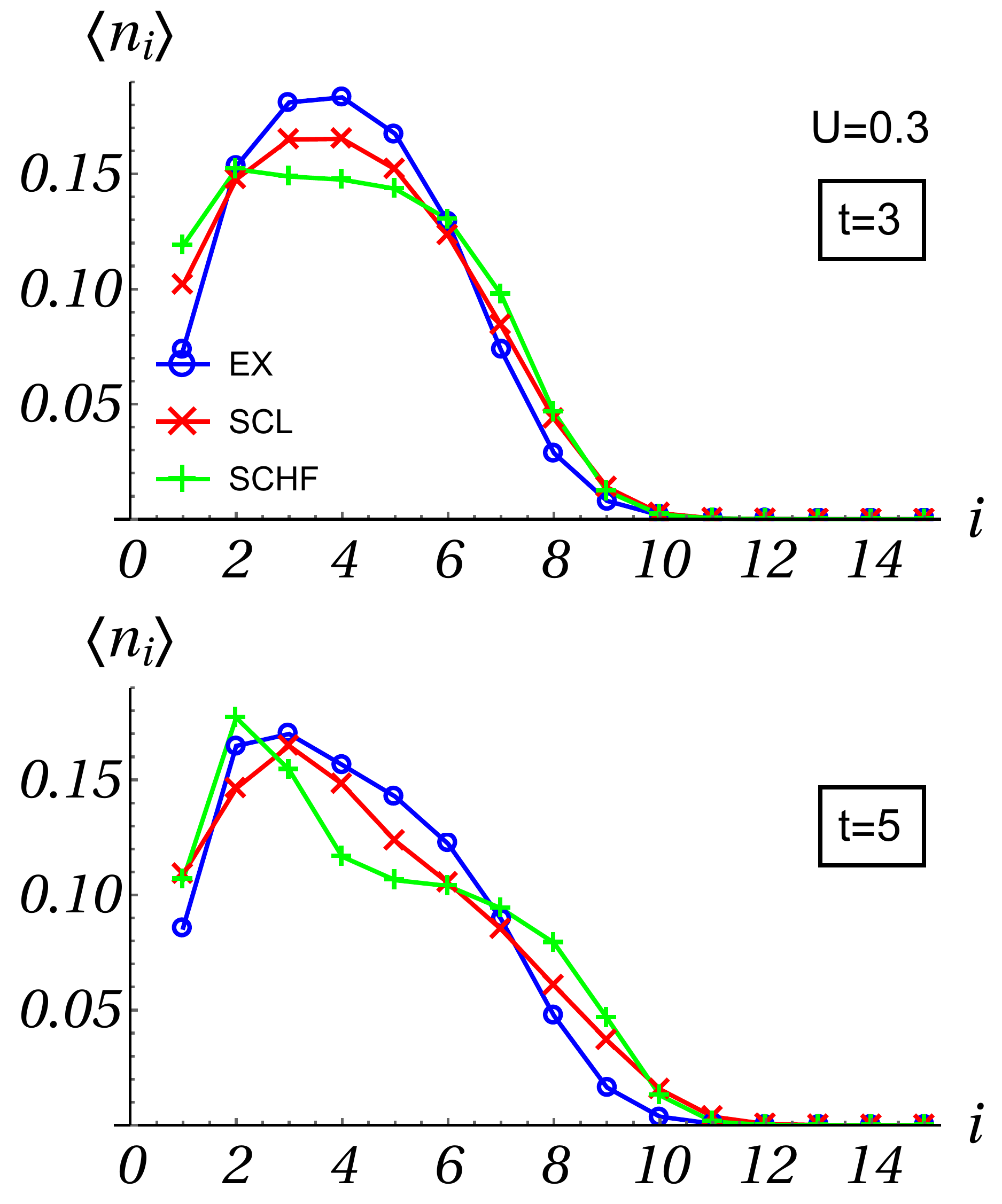}&\includegraphics[
width=4cm]{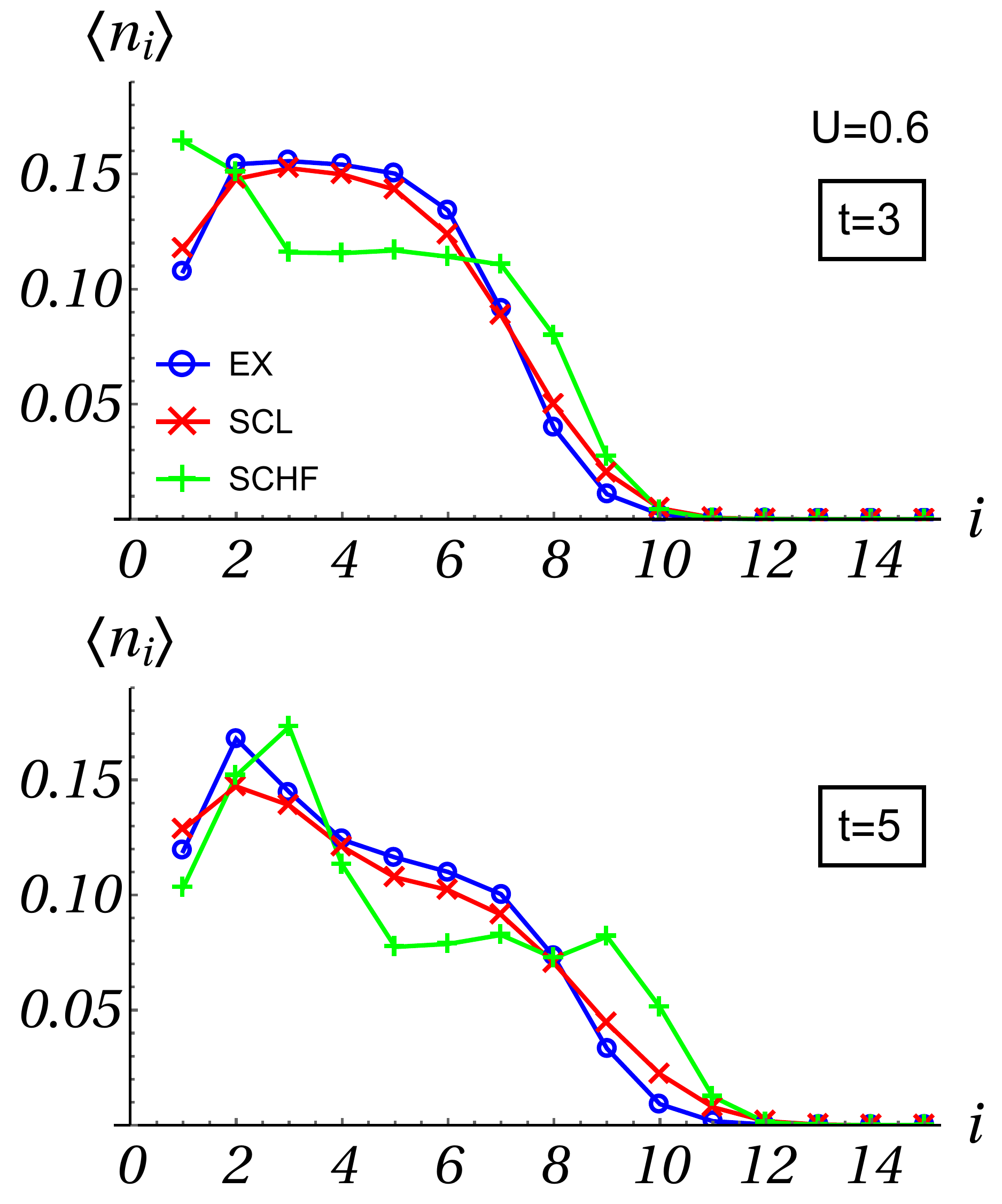}
\end{tabular}
 \caption{(Color online). Average number of bosons per site at two different 
times (upper row) $t=3$ and (lower row) $t=5$ and for two different final 
interactions (left column) $U=0.3$ and (right column) $U=0.6$. Here for an inhomogeneous system (see text) 
with $N=15$ sites, $n=5$ initially in its non-interacting ground state $U=0$.}
 \label{fig:numx-quad005-ns15}
\end{figure}

In Fig.\ref{fig:numx-ns15} and Fig.\ref{fig:numx-quad005-ns15} 
we show the average number of bosons
per site for a sistem with $N=15$ sites and $n=5$ bosons ($\rho=n/N=1/3$)
for the homogeneous and inhomogeneous case respectively.
We compare the exact dynamics (EX) (blue circles),
self-consistent ladder approximation (SCL) (red x's) and the self-consistent 
Hartree-Fock (SCHF) (green crosses).
In each figure, plots refer to two different times (top) $t=3$ and (bottom) $t=5$ 
and for two different values of the interaction: $U=0.3$ (left panel) and $U=0.6$ 
(right panel).
We can see that the SCL approximation captures really well the 
main features of the exact evolution both qualitatively and
quantitatively as opposite to the SFHF
which fails to describe both short and long time dynamics
(here long times is with respect to the perturbation reaching the boundaries).
Moreover we see that even at higher interactions the SCL still gives accurate results.
This might seem surprising at first sight but it is not,
the ladder approximation, by construction, includes contribution
from multiple particle-particle scattering. 
Together with the self-consistent approach this allows including these diagrams 
to all orders in $U$. The only limitation comes therefore 
from physical processes which are not encompassed by the ladder expansion.

\begin{figure}
\begin{tabular}{cc}
\includegraphics[width=4cm]{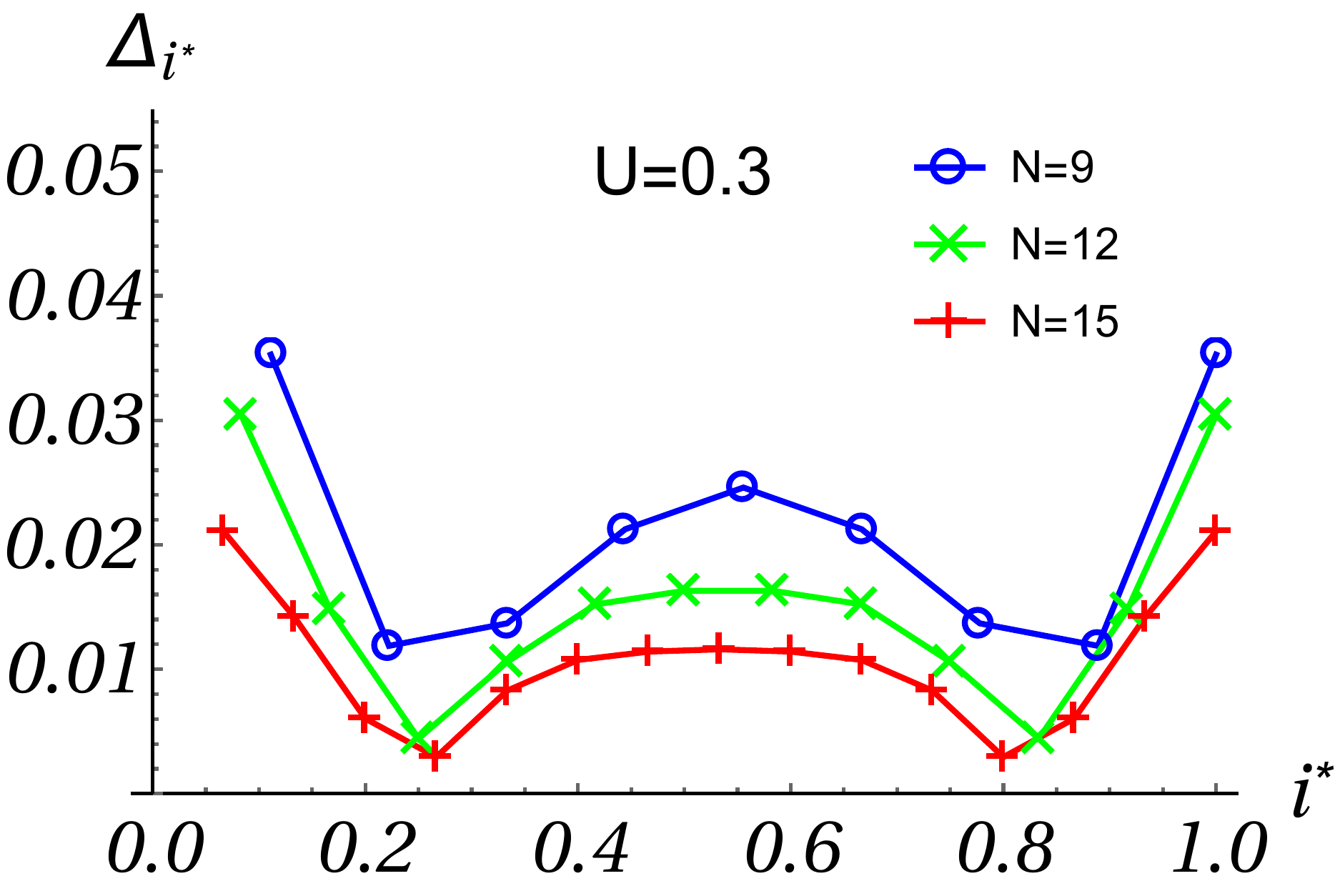}&\includegraphics[
width=4cm]{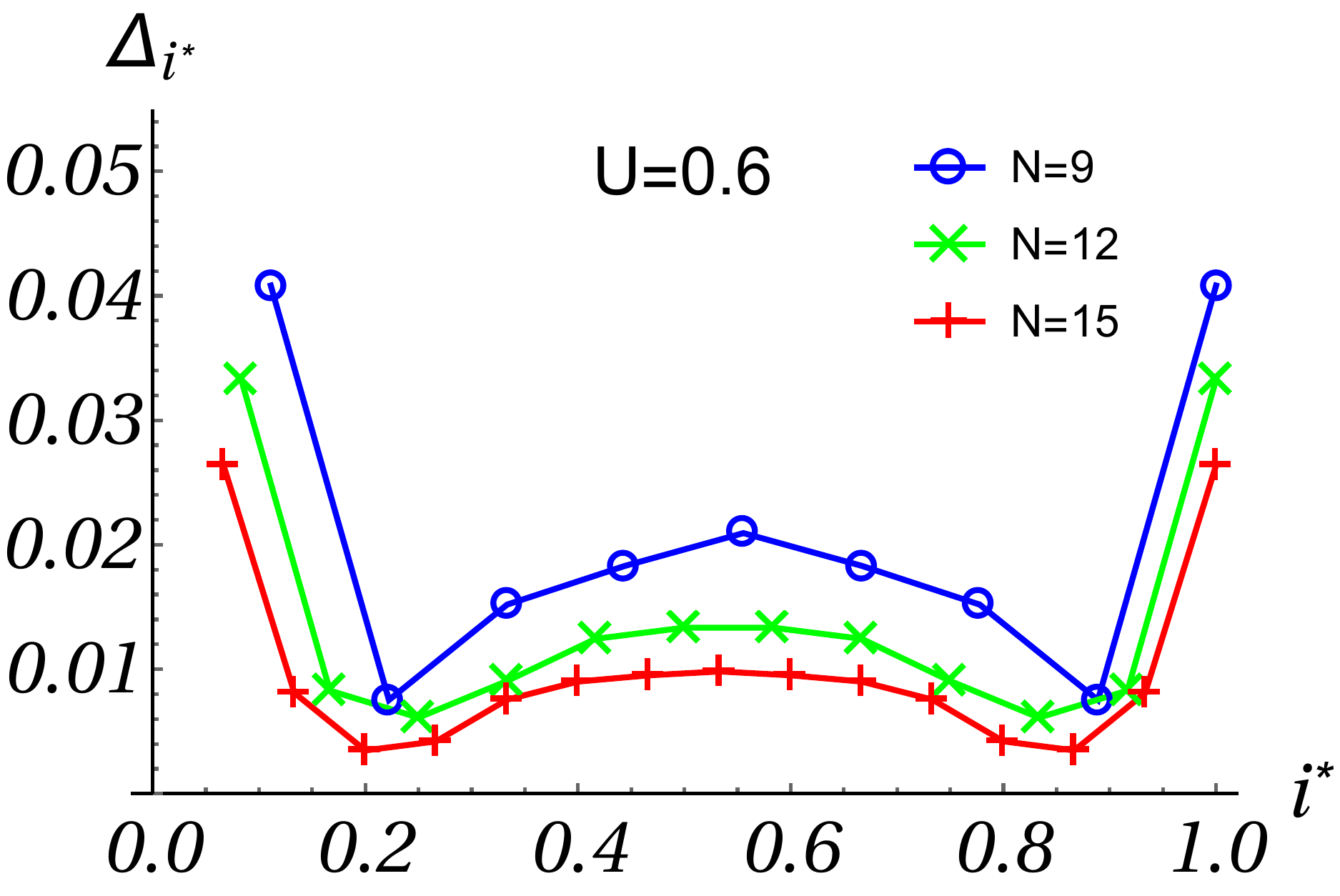}\\
\includegraphics[width=4cm]{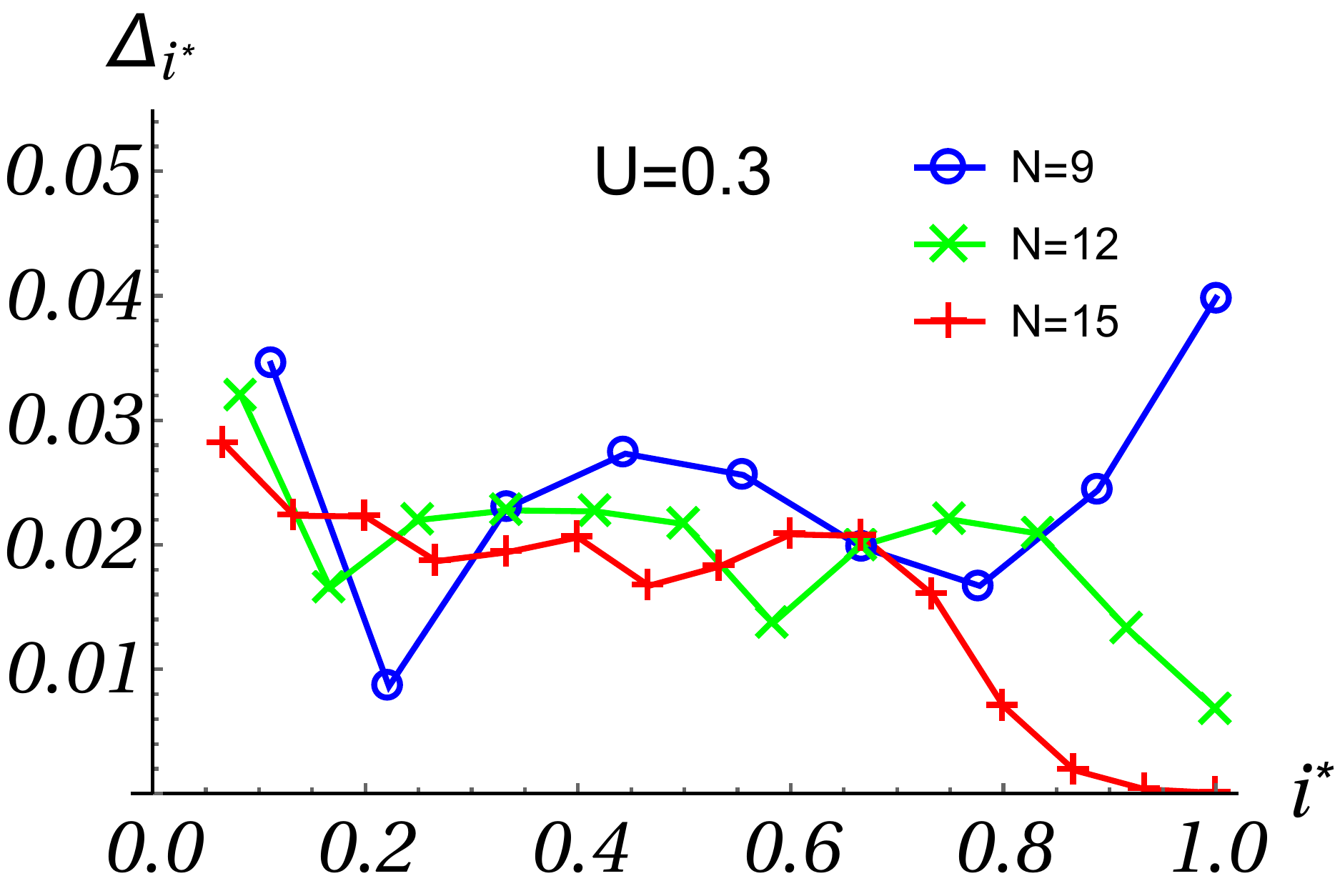}&\includegraphics[
width=4cm]{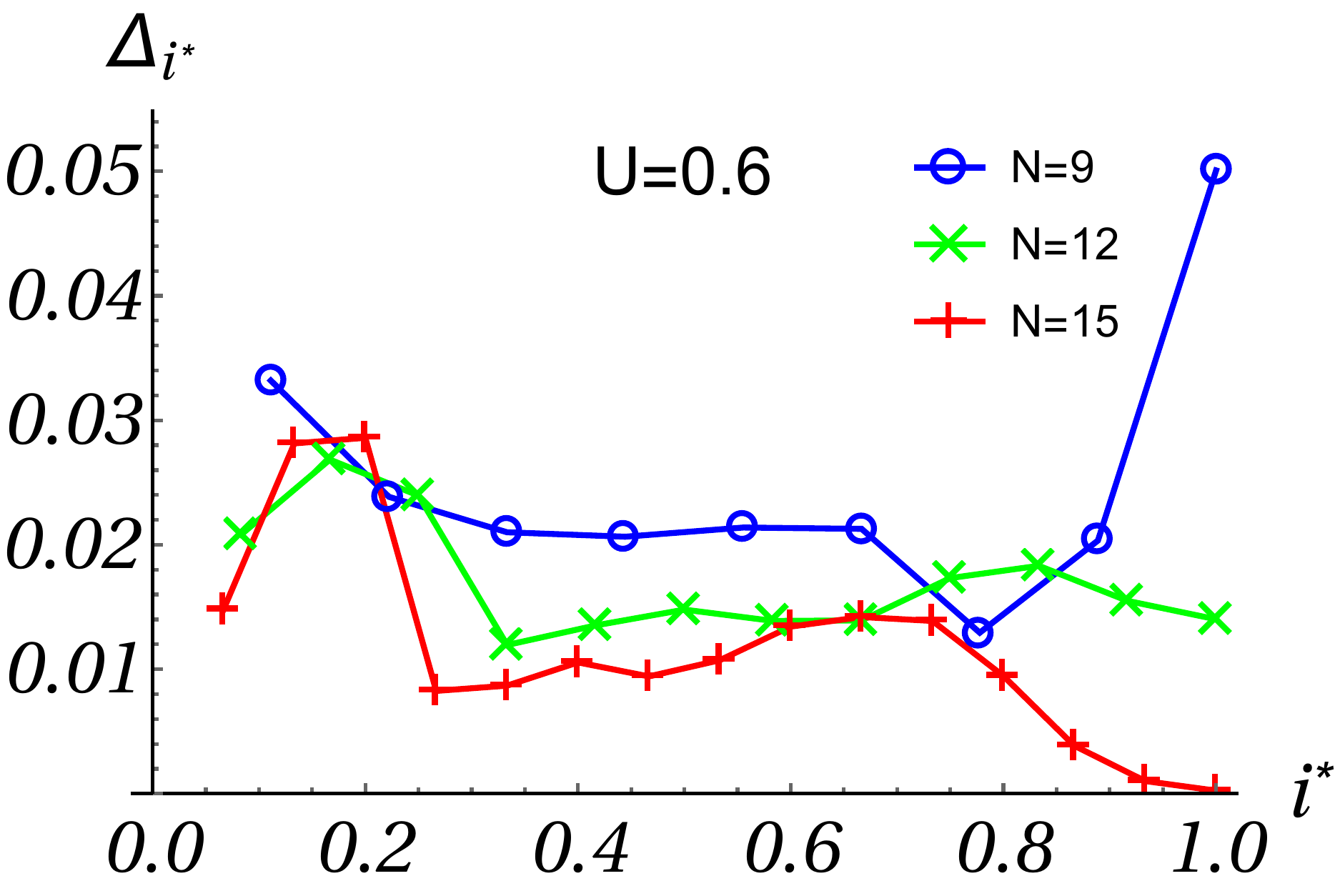}
\end{tabular}
 \caption{(Color online). Maximum deviation per site ($i^*=i/N$) over total time interval 
 beteewn the EX solution and the SCL for three different system sizes:
 (blue) $N=9$, (green) $N=12$ and (red) $N=15$.}
 \label{fig:errs}
\end{figure}

We can also show that,
in the regimes considered here (dilute gases and/or weak interactions),
the discrepancy between the EX and the SCL reduces with the system size.
In Fig.\ref{fig:errs} we plot the maximum deviation over the time
interval at each site $\Delta_i=max_t |n_i^{EX}(t)-n_i^{SCL}(t)|$
for three different system sizes: (blue) $N=9$, (green) $N=12$, (red) $N=15$.
Increasing the system size the discrepandy form the EX solution 
of the SCL decreases both for the homogeneous (top) and the inhomogeneous (bottom) case.

\begin{figure}
\begin{tabular}{c|c}
\includegraphics[
width=4.5cm]{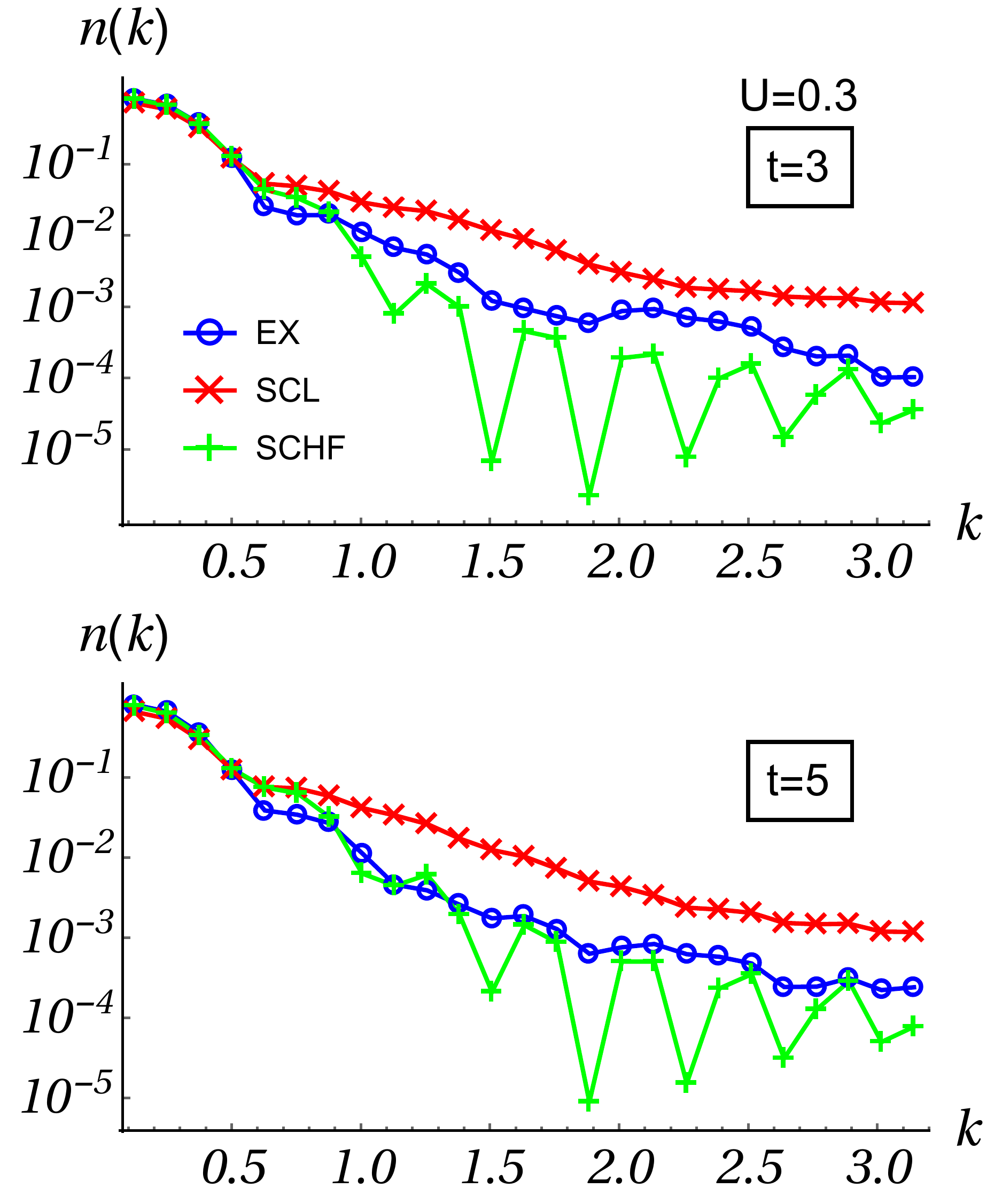}&\includegraphics[
width=4.5cm]{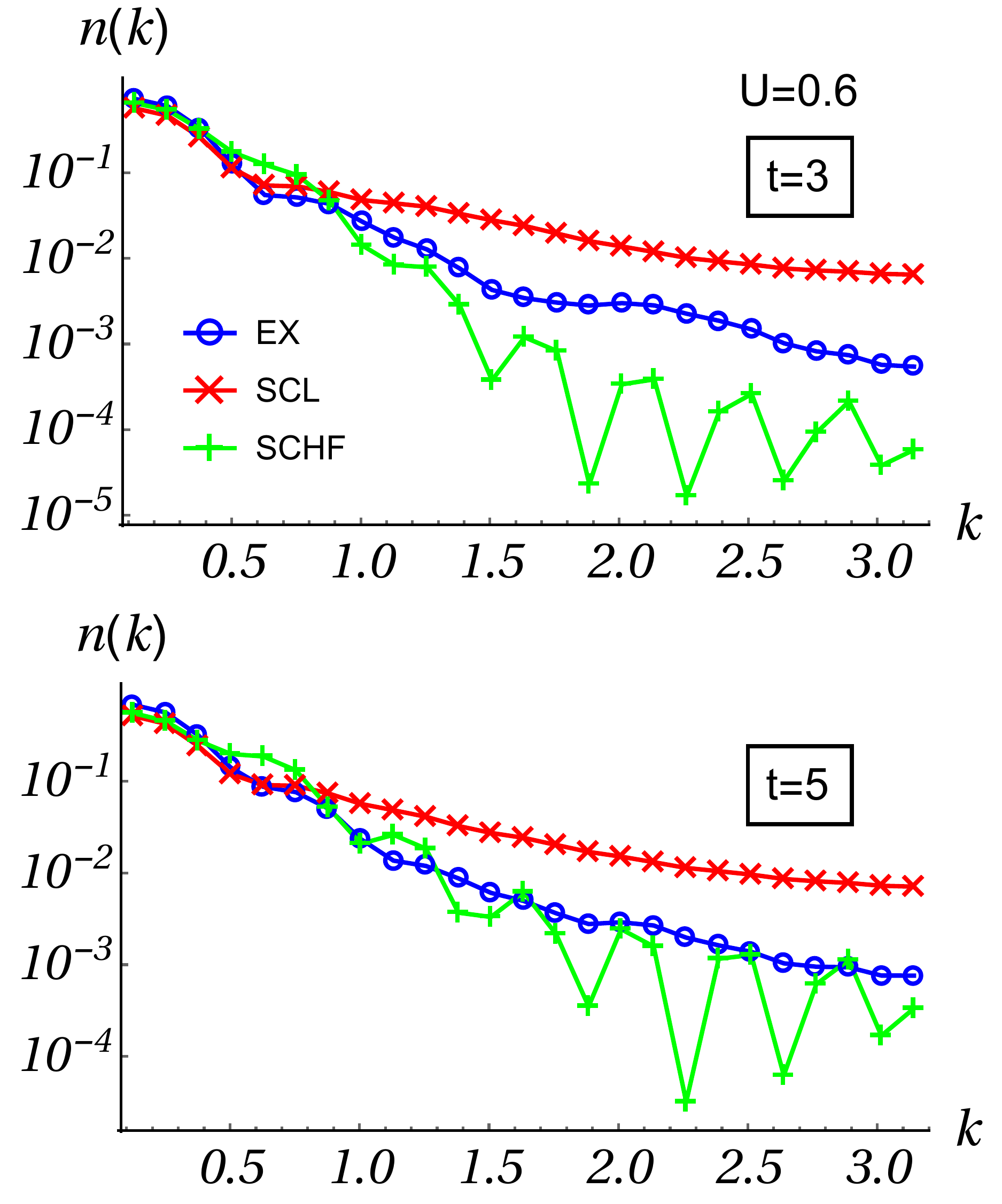}
\end{tabular}
\caption{(Color online). Momentum distribution for the homogenous case
 and for a system with $N=15$ sites and $n=5$ bosons at two different times: (top) $t=3$
 and (bottom) $t=5$ and for two different final interaction strenghts: (left) $U=0.3$
 and (right) $U=0.6$.}

 \label{fig:numk-ns15}
\end{figure}

\begin{figure}
\begin{tabular}{c|c}
\includegraphics[
width=4.5cm]{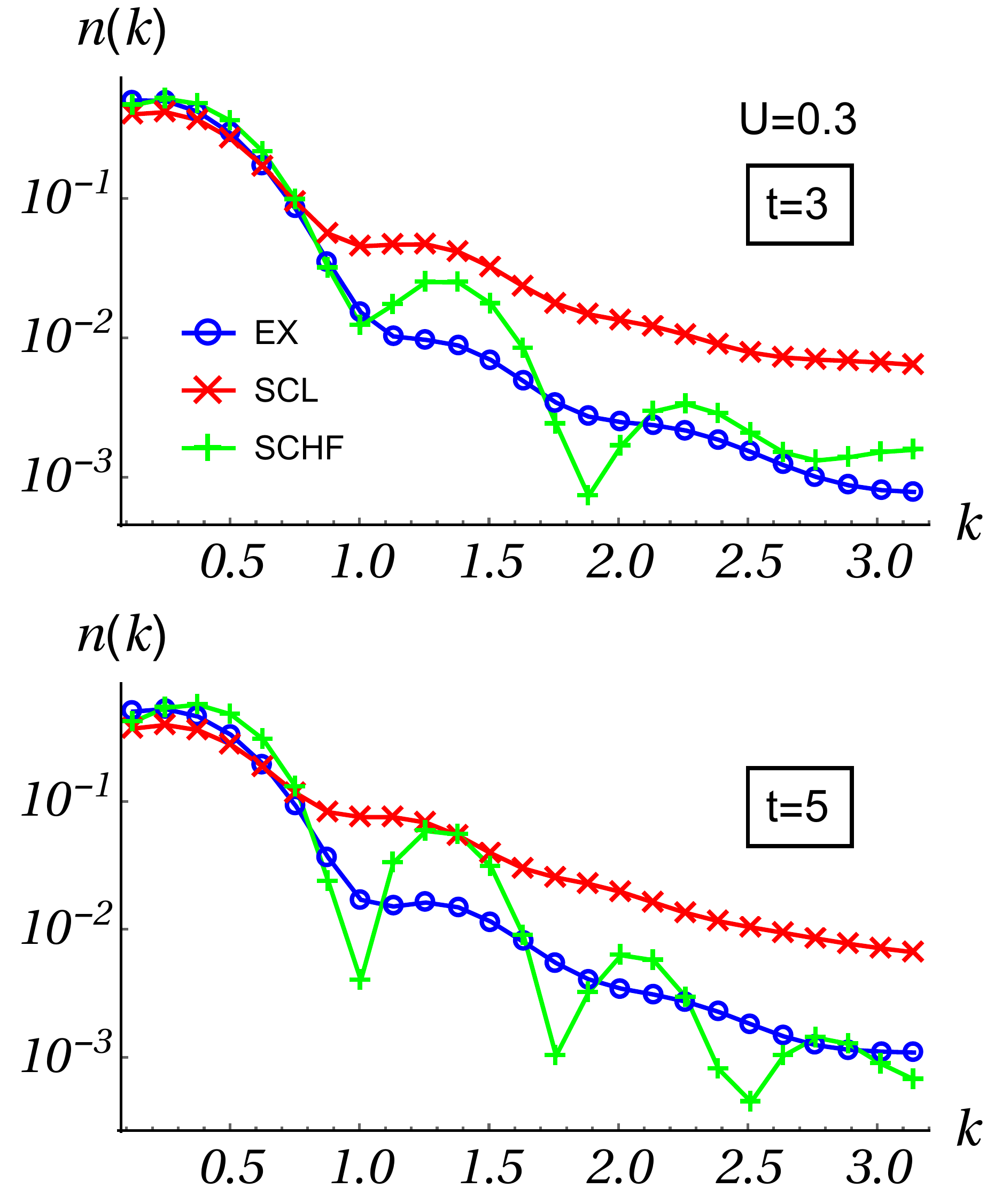}&\includegraphics[
width=4.5cm]{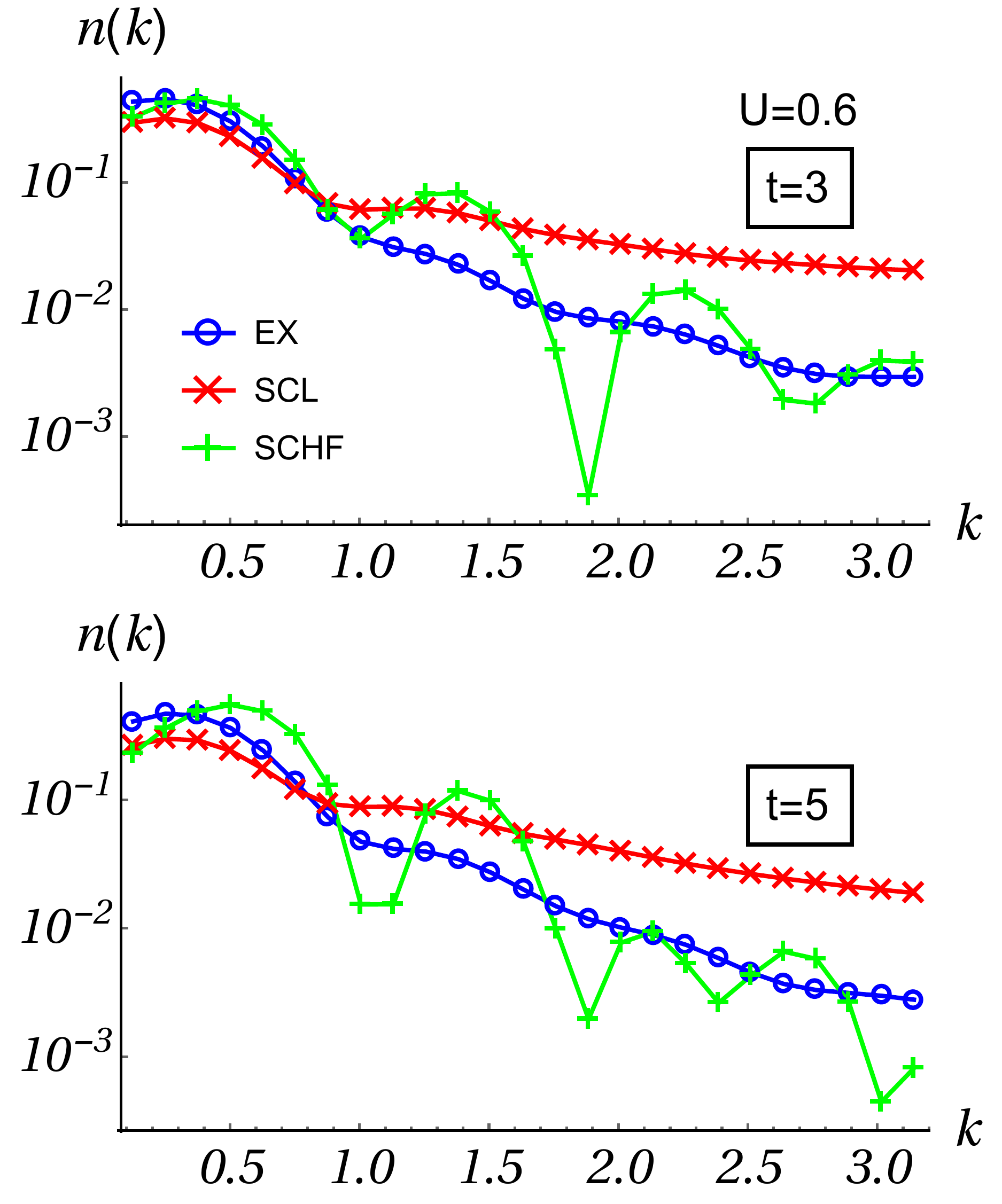}
\end{tabular}
 \caption{(Color online). Momentum distribution for the inhomogenous case (see text)
 and for a system with $N=15$ sites and $n=5$ bosons at two different times: (top) $t=3$
 and (bottom) $t=5$ and for two different final interaction strenghts: (left) $U=0.3$
 and (right) $U=0.6$.}
 \label{fig:numk-quad005-ns15}
\end{figure}

Beside the density profile it is worth mentionning that the SCL
also captures the main features of the equal-time correlation functions.
In Figs.\ref{fig:numk-ns15} and \ref{fig:numk-quad005-ns15}
we plot the momentum distribution obtained from the single particle density 
matrix $\mean{\opt{b}{t}{i}{\dag}\opt{b}{t}{j}{}}{0}$
for two different times and interaction strenghts.

It can be seen that the SCL follows the behavior of the EX
solution althoght showing a deviations at high momenta $k$.
It is interesting to observe that for the inhomogeneous case
the SCL shows the "plasmonic"-like excitation
around $k\approx 1.2$ which is manifested as a plateaux in $n(k)$.
This is nothing but a density wave-packet traveling towards the right boundary.

\begin{figure}
\begin{tabular}{cc}
\includegraphics[width=4cm]{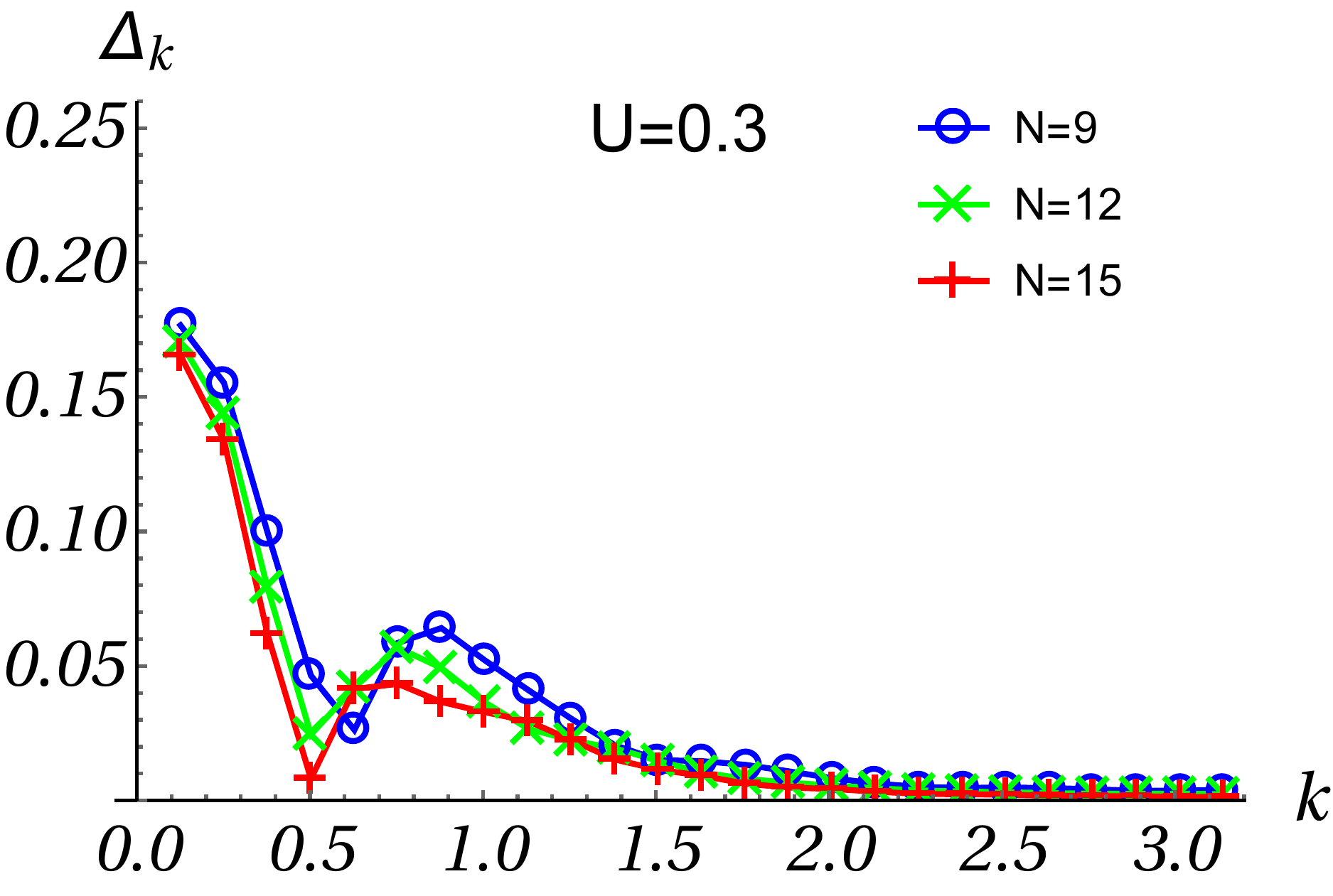}&\includegraphics[
width=4cm]{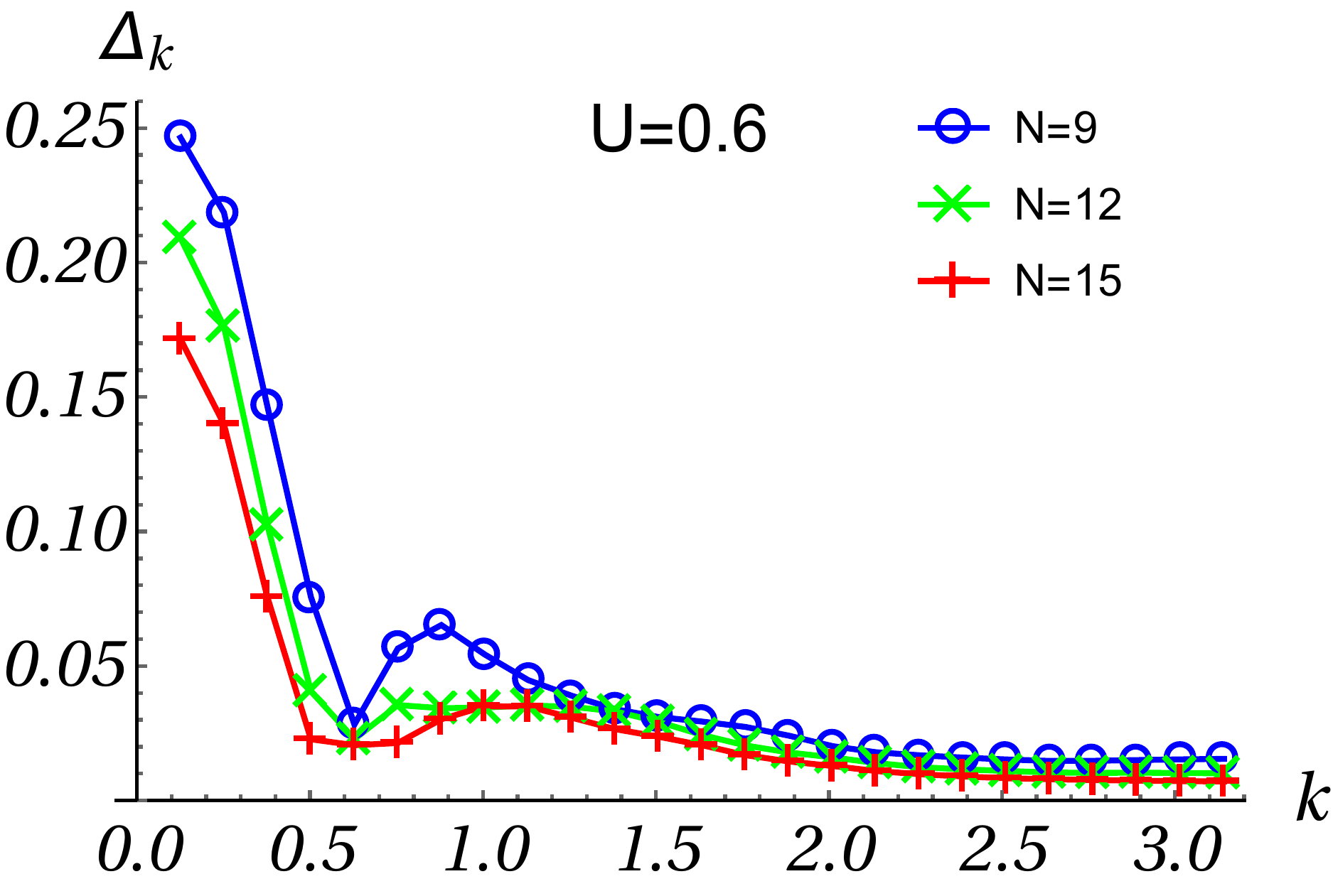}\\
\includegraphics[width=4cm]{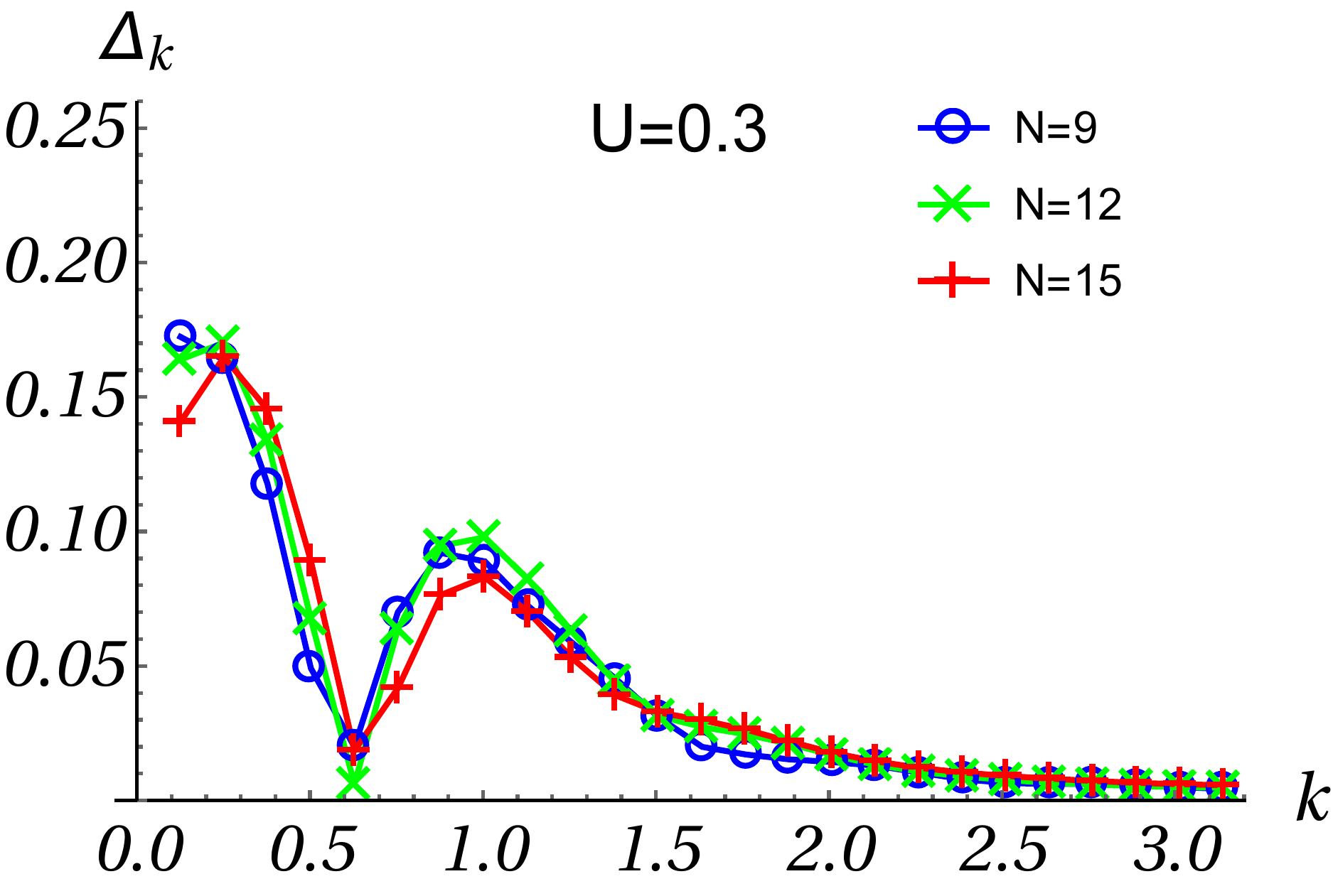}&\includegraphics[
width=4cm]{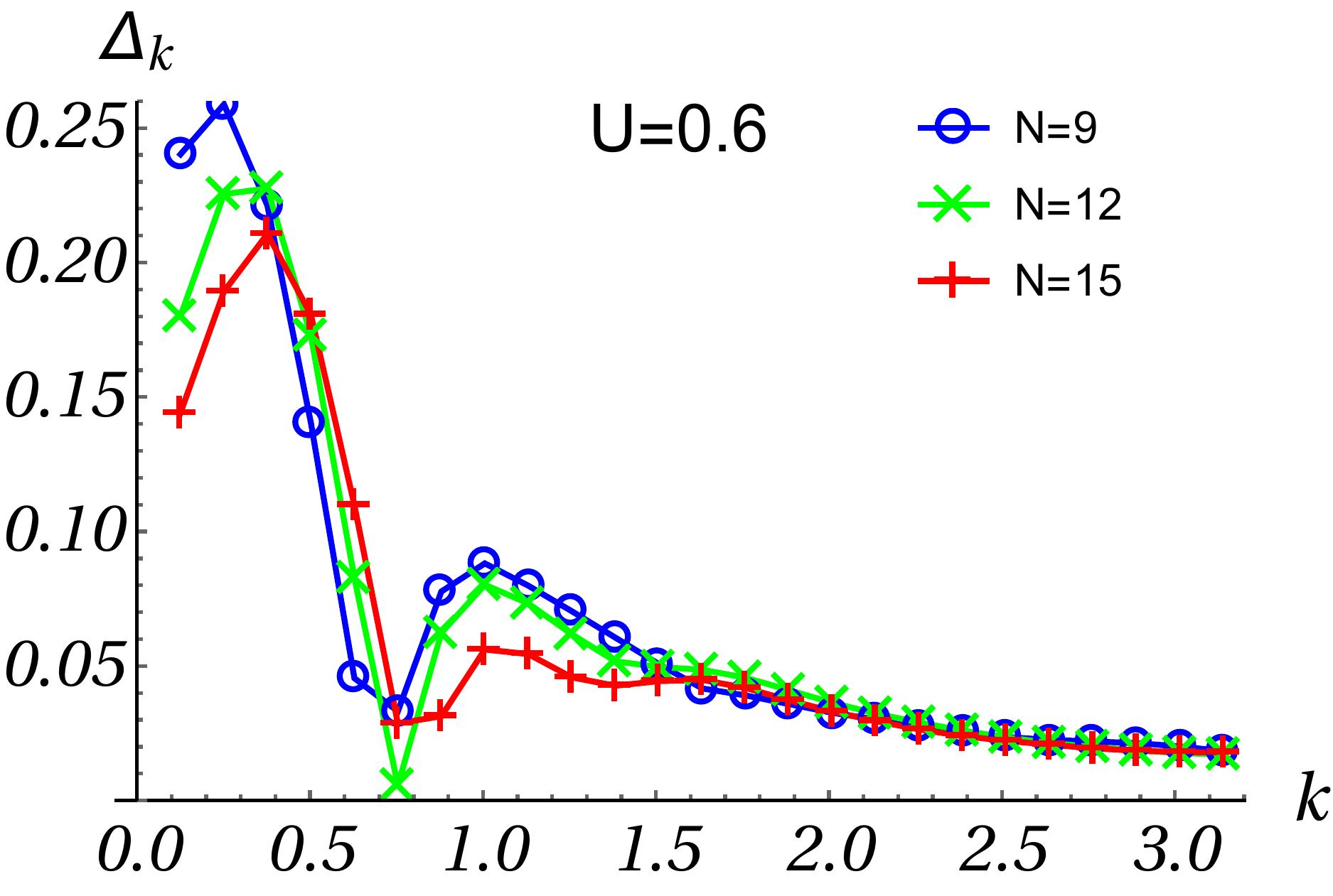}
\end{tabular}
 \caption{(Color online). Maximum deviation per momentum $k$ over total time interval 
 beteewn the EX solution and the SCL for three different system sizes:
 (blue) $N=9$, (green) $N=12$ and (red) $N=15$.}
 \label{fig:errsk}
\end{figure}

In Fig.\ref{fig:errsk} we show again the maximum deviation 
per momentum $k$ over the whole time evulution
for different system sizes.
It can be observed that the general behavior is that the maximum
deviation decreases with sistem size.

Summarizing we can say that 
the comparison of the self-consistent ladder with the exact diagonalization,
at zero temperature and for small systems, 
shows that we can rely on this approach to describe the dynamics of bigger systems
On the other hand, we can use the comparison at zero temperature to infer that 
there will be agreement also at finite temperature due to the fact that the total Hamiltonian
does not couple subspaces of the Hilbert space with different total number of bosons.
The dynamical quantities for an initial Gibbs state are the weighted average 
of quantities evolving in subspaces with fixed number of bosons.

\end{appendix}

\end{document}